# Entropic stabilization of proteins and its proteomic consequences.

Igor N. Berezovsky[1], William W. Chen[1,2], Paul J. Choi[1]

and Eugene I. Shakhnovich[1,*]

[1]Department of Chemistry and Chemical Biology and [2]Department of Biophysics,

Harvard University, 12 Oxford Street, Cambridge, MA 02138

*Correspondence should be directed to Prof. Eugene I. Shakhnovich, Department of Chemistry and Chemical Biology, Harvard University, 12 Oxford Street, Cambridge, MA 02138, phone: 617-495-4130, fax: 617-384-9228, e-mail: eugene@belok.harvard.edu

## Abstract

**Background.** Evolutionary traces of thermophilic adaptation are manifest, on whole genome level, in compositional biases towards certain types of aminoacids. However sometimes it is hard to discern their causes without clear understanding of underlying physical mechanisms of thermal stabilization of proteins. For example it is well-known that hyperthermophiles feature greater proportion of charged residues but, surprisingly, the excess of positively charged residues is almost entirely due to Lysines but not Arginines in the majority of hyperthermophilic genomes.

**Principal Findings** All-atom simulations show that Lysines have much greater number of accessible rotamers than equally buried Arginines in folded states of proteins. This finding suggests that Lysines would preferentially entropically stabilize the native state. Indeed we show in computational experiments that Arginine-to-Lysine amino acid substitutions result in noticeable stabilization of proteins. We then hypothesize that if evolution uses this physical mechanisms in its strategies of thermophilic adaptation then hyperthermostable organisms would have much greater content of Lysines in their proteomes than of comparable in size and similarly charged Arginines.. Consistent with that, high-throughput comparative analysis of complete proteomes shows extremely strong bias towards Arginine-to-Lysine replacement in hyperthermophilic organisms and overall much greater content of Lysines than Arginines in hyperthermophiles. This finding cannot be explained by genomic GC compositional biases.

**Significance** We discovered here a novel entropic mechanism of protein thermostability due to residual dynamics of rotamer isomerization in native state and demonstrated its immediate proteomic implications. Our study provides an example of how analysis of a fundamental physical mechanism of thermostability helps to resolve a puzzle in comparative genomics as to why aminoacid compositions of hyperthermophilic proteomes are significantly biased towards Lysines but not similarly charged Arginines.

## Introduction

Enhancing the stability of globular proteins remains an important task of protein engineering and design [1,2]. The major mechanisms for increasing stability discovered so far vary from introduction of additional chemical bonds (e.g. disulfide bridges) or ion pairs [3,4]to increasing either the enthalpic free energy contributions by optimizing of hydrophobic core interactions [5-9] or the entropic contributions by varying main-chain degrees of freedom in unfolded state [10]. This repertoire of mechanisms relies on a variety of underlying physical principles for increasing protein stability [11]. The diversity of extreme environments and the long evolutionary history of their organismal proteomes [12,13] suggest, in turn, many hypothetical mechanisms of protein stabilization in response to the demands of the environment. Furthermore, the fact that each proteome contains a variety of structures and functions suggests that nature used all, even seemingly negligible, opportunities and their combinations for structure stabilization when adapting to extreme environmental conditions [13]. Here we show how side-chain entropy in the native state can be a factor in thermophilic adaptation. The analysis of statistics of rotameric states, together with computational mutation experiments, followed by high-throughput analysis of complete proteomes reveals a previously unknown mechanism of stabilization via replacement of Arginine residues with Lysines. This substitution stabilizes the folded state, yet it preserves the charged nature of the substitution position, which may be important for other, perhaps functional, reasons. Thus, possible evolutionary advantages of this mechanism are as follows: (i) avoidance of sterically unfavorable contacts upon substitution, (ii) conservation of the similar-to-the-original (in terms of geometry and size) side chains, and (iii) preservation of the positive charge. These subtle advantages exemplify the elegant work of natural selection and hint at the existence of other, yet undiscovered, mechanisms of protein adaptation.

## Results

### Monte Carlo unfolding simulations of Hydrolases H from *E. coli* and *T. thermophilus*

The Gō model of protein folding is an idealized model in which the favorable interaction contact terms are exactly those found in the native structure [14]. In this model, the physico-chemical details of protein interactions are replaced by a generic contact energy term that is the same for all contacts between atoms that are found in contact in the native structure, though the complexity of the folded backbone and the side-chain conformations are preserved. It has been argued that such a model is a good representation of such aspects of protein energetics and folding where non-native contacts do not play a massive role [14,15]. It remains unclear whether such an idealized model can quantitatively predict absolute folding transition temperatures. However, our results suggest that the Gō model predicts the transition temperature accurately enough to discriminate between proteins of thermophilic and mesophilic origin.

Figure 1 shows the unfolding curves for the pair of meso/thermophilic Hydrolase H from *E. coli* (PDB code 1INO) and *T. thermophilus* (PDB code 2PRD). The Gō model correctly predicts a slightly higher transition temperature for the protein from thermophilic *T. thermophilus* compared to the one from *E. coli* (Figure 1). Remarkably, we observe the two unfolding curves coincide up to the transition, at which point they separate and then recombine at higher temperatures (Figure 1). Because the native states are enthalpically identical, and the folds are essentially the same, we surmise the origin of the transition temperature difference to be purely entropic. Specifically, given the nature of the Gō model, the entropic differences must arise from the different number of accessible rotamer states in different proteins. Calculation of average number of rotamers per residue in fully unfolded state [16] gives values of 12.0 and 11.4 for the mesophilic and the thermophilic proteins, respectively. These numbers thus demonstrate that the higher side-chain entropy in the unfolded state of mesophilic Hydrolase is partially responsible for the fact that it unfolds at a lower temperature than the thermophilic structure.

**Lysine and Arginine: Archetypal signal of rotamer entropy in Gō unfolding**

A careful look at the number of accessible states for each residue type in the folded state of Hydrolases (Table 1) lead us to another interesting observation: although Arginine and Lysine are chemically similar and have the same *maximal* number of possible rotameric states -81- they differ greatly in their rotameric accessibilities in the folded state.

There is total of five groups of amino acid residues with the same maximal number of rotamers in unfolded state (see also Table 1): (i) Arg, Lys (maximal number of rotamers is 81); (ii) Glu, Met (27); (iii) Ile, Leu, His, Trp (9); (iv) Asp, Phe, Tyr (6); (v) Cys, Ser, Thr, Val (3). In both proteins the average number of rotamers per residue in the folded state is significantly greater for Lys residues than for Arg residues: 20.1 and 17.2 versus 3.5 and 5.6 rotamers per residue in the folded state, respectively. For the other residues with the same or similar physical and chemical features and the same maximal number of rotameric states (pairs Leu/Ile, Thr/Ser, Val/Thr, and Phe/Tyr), the difference in the number of rotamers in the folded state is rather negligible (Table 1). This suggests that Lysine and Arginine provide an excellent platform to test possible entropy-based mechanism of protein stabilization for both genomic and computational studies, since Arg/Lys substitutions (1) preserve enthalpic contributions, (2) maintain the same physical and chemical features, and (3) give differing entropic contributions (Table 1). Accordingly, we study the effects of side-chain entropy on protein stability for the chosen pair of types of amino acid residues, Arginine and Lysine [10,17-21].

**Lysine and Arginine: Possible Archetypal Signal of Rotamer Entropy in Protein Stability**

The amino acids Lysine and Arginine are both positively charged; both contain at least five heavy atoms in their side chains. Both amino acids have 4 degrees of freedom in their rotatable bonds. The guanidinium group at the end of Arginine displays resonance, and, as a consequence, has no internal rotational freedom. The salient difference between Arginine and Lysine is the fact that Lysine is less bulky. Therefore, in the folded state, Lysine may have slightly more freedom. Estimates of solvent accessibility of Arginine and Lysine do not reveal a substantial difference (data not shown). As a control comparison, we use the pair Isoleucine/Leucine (each residue has a maximum of 9 rotameric states and similar to each other physical and chemical properties). Using the Gō model, we sample the number of accessible rotamers as a function of temperature for the Hydrolases from *E. coli* and *T. thermophilus* (Figure 2). We approximate the entropy of the side-chain with the natural logarithm of the number of observed states in long equilibrium Monte-Carlo simulation. In the case of Lysine, there is substantial residual entropy in both the folded state and the unfolded state (Table 1). This is in contrast to Arginine, which shows a marked decrease of rotamer freedom in the folded state making Arginine less free energetically favorable than Lysine for native state stabilization. The control pair Isoleucine/Leucine shows no such difference in the number of rotamers in the folded state (Table 1).

These results suggest that Arginine and Lysine are an excellent candidate pair to use as a test platform for both genomic and computational studies, for two reasons: (1) they have similar physico-chemical properties and thus preserve enthalpic interactions upon mutation from one to the other, and (2) they have similar rotamer entropies in unfolded state but different rotamer entropies in the folded states.

Figure 2 shows the temperature dependence of the natural logarithm of the number of rotamers for pairs Arg/Lys (charts a, b), Leu/Ile (c, d), Thr/Ser (e, f), and Thr/Val (g, h) in Hydrolases H from *E. coli* (1INO) and *T. thermophilis* (2PRD). According to Table 1, Lysine and Arginine residues have differentl residual side chain entropy: Lysine residues have many more rotamers in the folded state than Arginine residues: in average, 20.1 and 17.2 versus 3.5 and 5.6 rotamers per residue of a particular type (Lys or Arg) in 1INO and 2PRD, respectively. The control group in this analysis is the pair Leu/Ile, which show a highly similar temperature dependence of the number of rotamers (Figure 2 c, d) for both proteins. Two last pairs Thr/Ser and Thr/Val confirm the role of the side-chain size and, as a consequence, its flexibility in providing number of accessible rotameric states. Each of, Thr, Ser, and Val have a maximum of 3 possible rotamers and, thus, can be also compared. Although both Thr and Ser are hydrophilic residues, Ser residues have a slightly higher number of rotameric states in the folded structure (at absolute temperature 1 from MC simulations) as a result of its smaller side-chain. The hydrophilic/hydrophobic pair Thr/Val (Figure 2 g, h) exhibit very similar behavior stemming from the similarity of their side-chain geometries. This result is further substantiated by the temperature dependence data for the pairs Val/Ser and Phe/Tyr. ( The results for averaged temperature dependence, for residue types from both Hydrolases, are presented in Supplementary Materials, Figure S1 a-f). The bulky side-chains of both Phe and Tyr (Figure S1 f) show practically the same temperature dependence, whereas in the pair Val/Ser (Figure S1 e) the latter has slightly more rotamers in the folded state.

***Statistics of rotameric states in a representative set of protein structures***

Let us consider situation when residues with similar physical and chemical properties have a different number of rotamers in the folded state. The similarity of physical and chemical properties makes it possible to adjust stability due to entropic factor by mutating one residue type into another without changing the structure significantly. The first step to verify this mechanism is a statistical study of the difference in number of accessible rotamers for the folded and unfolded states. We analyzed the ratio of the number of rotamers (in natural logarithm units) at absolute temperature T=4 (completely unfolded state) to that at T=1 (folded state) for a representative set of 18 protein structures. Figure 3 shows histograms of ratios for the following pairs of aminoacid residues: Arg/Lys (chart a), Val/Thr (b), and Phe/Tyr (c).

Arginine and Lysine show significantly different rotamer number ratios in the folded state distribution (Figure 3a; mean values of the distribution for Lysine and Arginin are 2.14 and 1.21 in natural logarithm units, respectively). Ratios for the pairs Leu/Ile, Val/Thr, and Phe/Tyr are very similar (Figure 3b-d), with mean values of distributions are 1.7/1.4, 0.85/0.97, and 1.62/1.58, respectively. This data corroborate, that Lysine residues contribute entropically to the change in equilibrium between the unfolded and folded states, whereas residues in pairs Leu/Ile, Val/Thr, and Phe/Tyr have similar number of rotamers in the folded state. As a next step we prove the stabilizing role of Lysine vs Arginine in direct computer simulation experiment.

***R/K replacement computational experiment. Detecting changes in stability by Monte Carlo unfolding simulations***

As stated above, both a statistical analysis of rotameric states and a comparative high-throughput analysis of complete proteomes demonstrate the particular role of Lysine rotamers in protein thermostability. To demonstrate the stabilizing role of Lysine residues we make a replacement of type Arg/Lys and analyze the unfolding simulations in order to detect an anticipated increase in structure stability.

We replaced Arginine residues with Lysine residues in corresponding positions and locally minimized the resulting structures. We left the rest of the structure intact (see Materials and Methods) in order to influence the native structure as little as possible. The same local minimization was applied to the native structure.

Results of the replacements of Arginine residues with Lysine residues in both Hydrolases H are as follows: in Hydrolase H from *E. coli* position 43 (20 Arg-residue rotamers / 31 Lys-residue rotamers in the folded state), 86 (2/5), 88 (4/12), and 171 (3/18); in Hydrolase H from *T. thermophilus* position 24 (1/5), 43 (4/12), 114 (15/47), 156 (14/24), 158 (5/12), 166 (21/25), and 171 (3/10). We also analyzed combinations of Arg-to-Lys replacements in different positions in the structure. We found an increase of transition temperature in the replacement R171K and in combination of all R/K substitutions in positions 43, 86, 88, and 171 in mesophilic Hydorlase from *E. coli*. Figure 4 (a, b) shows a plot of the temperature dependence of the energy in unfolding simulations of structures with replacement R171K in Hydrolases H from *E.coli* and *T. thermophilus* [22-24]. Though there is a slight increase in the enthalpic term in the modified (R171K) structure of thermophilic Hydrolase (3321 native contacts in the modified structure versus 3246 in the original, according to the Gō model), and an increase in the number of rotamers in the modified structure, there is no indication of a change in the transition temperature in unfolding simulations (Figure 4a). Similar replacements in the structure of mesophilic Hydrolase H from *E. coli*, on the other hands, cause a change in the transition temperature of approximately 0.1 in absolute units (2.6 percent). The increase in the number of native contacts in the modified structure (3226 in modified versus 3131 in original) accounts for 3 percent of the difference in transition temperature; entropic factors do not play a stabilizing role in this case (5.38 and 5.34 rotameric states per residue in original un mutated structures, respectively). We detected an increase in the stability of the structure when all Arginine residues (positions 43, 86, 88, and 171) were replaced by Lysine residues. Taking into account both the decrease in the enthalpic term in the modified structure (3083 native contacts versus 3102 in the original, or approximately 0.6 percent loss) and the simultaneous increase in the transition temperature by 0.05 of absolute units (gain of 1.3 percent) gives a total increase of 2 percent in stability, which we conclude to be an effect of entropy stabilization of the structure. The number of rotamers per residue increases from 5.41 in the original to 5.62 in the mutated structure, a 4 per cent difference, which, taking into account the roughness of the estimate, corroborates an increase in stability. The absence of a stabilizing effect of replacements in thermophilic Hydrolase H from *T. thermophilus* can be explained by the high stability of the original protein, as demonstrated earlier [22-24].

We performed a similar experiment with a smaller protein to improve sampling. Cytochrome C from *R. sphaeroides* [25] contains 112 amino acid residues, with positions 24, 26, 53, 58, 74, 80, 87, and 95 occupied by Arginine residues. Simulations reveal the following variations in the number of rotamers in each position upon replacement with Lys: position 24 (8/16), 26 (15/17), 53 (37/38), 58 (5/22), 74 (2/6), 80 (43/43), and 95 (8/52). Replacement in individual positions did not reveal an increase in stability. However, simultaneous substitution of all Arginine residues by Lysines led to a noticeable increase in transition temperature, while the enthalpic term decreased by 0.5 percent (1607 native contacts in the modified structure compared to 1615 in the original). Figure 5 shows the temperature dependence of the energies, averaged over five runs (each $5x10^7$ MC steps). The difference between the transition temperature of the original and the modified structures is ΔT=0.07 absolute units (3 per cent) increase, which translates into a 3.5 per cent increase in stability when the unfavorable change in entthalpy is taken into account. Indeed, the mutated structure demonstrates an increase in the entropy of the folded state, 5.03 versus 4.56 rotameric states per residue in the original structure.

This data shows, that Lysine residues contribute greatly to the stabilization of folded states of proteins, compared to their peer positively charged Arginine , whereas residues in pairs Leu/Ile, Val/Thr, and Phe/Tyr have similar number of rotamers in the folded state. It is possible that this mechanism of stabilization is employed by Nature in its strategies of thermophilic adaptation. If this is the case it should be manifest in comparative genomics analysis in greater content of Lysines in hyperthermophiles compared to mesophiles and, importantly in bias towards Arg to Lys substitutions from mesophiles to hyperthermophiles.

## Analysis of complete proteomes

### *Amino acid composition biases in hyperthermophilic proteomes*

We performed quantitative analysis on 38 mesophilic and 12 hyperthermophilic proteomes. (For a list of the genomes used see Supplementary Materials, Tables S2, S3). We intentionally considered only hyperthermophilic proteomes in order to capture the most pronounced sequence biases associated with the extreme thermal stability of hyperthermophilic species. Figure 6 and Figure S2 show sets of composition histograms for two types of residues charged and hydrophilic, respectively, presumably associated with variations in thermal stability. While in thermophilic species, the percentage of polar residues is high [26], this percentage is the same or even smaller in hyperthermophilic organisms (for instance, Glu, Ser, Thr; see Supplementary Materials, Figure S2). In the case of charged residues we observe clear under-representation of Asp and His and an increase of Glu (Figure 6) in hyperthermophilic organisms. Increase of the Glu content is usually explained by its longer side-chain, which provides more opportunities for ion interaction [27,28]. In addition to the earlier-detected increase of total content in the Arg/Lys pair [27], we found that in 10 of the 12 hyperthermophilic genomes Lysine content is much higher (not less than 6 percent), whereas Arginine content is distributed evenly mostly between 3 and 6 percent (Figure 6 a,b). The dominance of Arginine in the pair Arg/Lys in proteomes of *M. kandleri* and *A. pernix* is an exception due to the high GC-content in these genomes[29,30]. Mean values for the percentage of Arg, Lys, His, Asp, and Glu in mesophilic and hyperthermophilic organisms (excluding M. *kandleri and A. pernix*), along with *p*-values according to binomial distribution calculated for the pair of archetypal representatives of each group, *E. coli* and *P. furiosus* (see Table S4 of Supplementary Materials): Arg, mesophilic/hyperthermophilic genomes, 5.46/4.94 ($p=8\cdot10^{-11}$); Lys, 5.35/8.48 ($p<10^{-14}$); Asp, 5.28/4.72 ($p<10^{-14}$); Glu, 6.1/8.42 ($p<10^{-14}$). Thus, 10 of the 12 hyperthermophilic organisms show difference in the preference for charged residues in mesophilic and hyperthermophilic genomes. (Note that for Arg and Asp the difference is inverse: there are *more* such groups in mesophiles than in thermophiles!) In particular, we detected an increase of Lysine content at the expense of Arginine content.

***Comparative analysis of hyperthermophilic versus mesophilic proteomes***

A persistently high percentage of Arg+Lys, though biased in most of the proteomes towards increased Lysine content, along with the similarity in physical and chemical features of these residues suggests an examination of substitutions of types R/K versus K/R in the alignment of mesophilic sequences (here, *E. coli*) versus hyperthermophilic ones. We started from the following hypothesis: if, as stated elsewhere [27], only the total content of Arginine plus Lysine residues matters in determining the stability of hyperthermostable proteins, then there should be no preference for one of the residues (Lys) over the other one (Arg). We used sequences of four hyperthermophilic archaea (*M. jannaschii, N. equitans*, *P. furiosis,* and *S. tokodaii*) and two hyperthermophilic bacteria (*A. aeolicus*, and *A. pernix*). Outputs of BLAST alignments were used for comparison of sequence substitutions that favor one or the other residue in each pair (see Table 2). Our data is presented in Table 2. The number before the slash is the percentage of amino acid residues in the mesophilic sequence, *e.g.* Leu that was replaced by the other amino acid in the hyperthermophilic sequence, *e.g.* Ile. The number after the slash reflects the same data for the opposite replacement, *e.g.* Ile, in the mesophilic sequence by Leu in the hyperthermophilic sequence. The control groups here are the pairs Leu/Ile and Ser/Thr; both residues in each pair are hydrophobic or polar, and both have the same maximal number of possible rotamers, 9 and 3, respectively. In all alignments of *E. coli* sequences against those from one of the hyperthermophilic genomes we obtained equal or very similar numbers of residues substitutions (numbers in parenthesis show ratio of forward to back substitutions). The exceptions are pairs LI/IL and RK/KR in *A. pernix*, which show bias in the opposite direction explained by the GC-content. Unlike the above control groups, the pairs RK/KR demonstrate a remarkable bias toward replacement of Arginine in the mesophilic sequence with Lysine in the hyperthermophilic sequence (at least 1.6 times in *P. furiosis*, and up to almost four times in *N. equitans*). P-values (calculated according to $\chi^2$ criteria) show a statistically significant preference for the Arginine-to-Lysine substitution as opposed to the reverse one. This challenges the idea that Arginine and Lysine play the same role in thermostability [27]. Therefore, the our comparative genomics analysis strongly supports the conclusion that Lysine has a particular or even exceptional role in protein stabilization [28].

## Discussion

### *Thermodynamical models of protein stability and the enthalpy/entropy relationship as a manifestation of a variety of stabilizing factors*

Most of the data on structure thermostability and its major factors comes from experiments aimed at analyzing the role of individual contributors, such as hydrophobic, van der Waals, electrostatic and other physical forces [11,12]. This determines a common computational approach to the analysis of protein thermostability: a limited dynamic or static model with a detailed Hamiltonian that partitions the forces into distinct classes [19,31,32].

The approach we presented here straddles the way between a complete description of folding and the limited dynamic models presented in previous studies. We employ a Gō model [14,15], which enables us to account for the enthalpically relevant terms, albeit in a coarse-grained manner. The Gō model also permits us to account accurately for the various entropic contributors to the folding free energy, namely the backbone entropy and side-chain entropies. Finally and most importantly, the simplicity of the model means that we are able to probe these various free energy effects with multiple folding runs relatively easily. In short, this approach makes it possible to examine the generic aspects of thermodynamics of thermostability.

Our results show the utility of Monte Carlo unfolding simulations with the Gō model as a way to detect the relative contributions of the free energy components, enthalpy and entropy. Furthermore, the description of the unfolding simulation in terms of the enthalpy/entropy relationship highlights the differences in the contribution of different types of amino acid residues to the entropic part of the free energy balance of a protein (see Table 1). We found a difference in the number of accessible rotamers *in folded state* despite the fact that these residues were naively expected to be fully fixed in native states i.e all have only one rotamer available in the native state. Logarithm of the ratio of the number of rotamers in the folded and unfolded states gives us the entropy difference upon folding for each residue (Figures 2 and 3). This data demonstrates significantly higher entropy of Lysine residues in folded states compared to those of Arginine.

We demonstrate here that our top-down approach, from analysis of thermodynamic quantities to discovery of concrete physical processes that give rise to the observed thermodynamic phenomena can not only detect differences in the free energies of stabilitization, but also reveal novel mechanisms of stabilization via the rotamer entropic effect.

***Genomic motivation for the novel mechanism of thermostability***

To validate a model of protein stability on the genomic/proteomic level, it is important to find particular expected compositional and sequence biases by means of massive high-throughput analysis. Even if the bulk of the protein in the organism exhibits a particular mechanism of stabilization according to the mechanism of adaptation commonly developed in the proteome [33-35], one or a few proteins may rely on a different/additional mechanism developed under specific environmental conditions. What additional information can we glean from the proteome analysis? First, amino acid compositional analysis reveals a bias toward Lysine residues in the pair Arg/Lys, typical for the genomes of hyperthermophiles. Such analysis demonstrates also a bias towards Lys and Glu in hyperthermophilic proteomes, whereas Asp and His are unfavorable in these organisms. The only exception are two hyperthermophilic genomes, *A. pernix* and *M. kandlerii*, whose preference for Arginine residues is a direct consequence high GC-content [28,30,36]. Second, comparative analysis of hyperthermophilic and mesophilic (here, *E. coli*) proteomes reveals an enrichment of Lysine content at expense of the Arginine.

Though bias in amino acid composition toward increasing charged residues is well documented in earlier works [26-28,37-40], the difference in the frequencies of Arginine and Lysine residues has not been explained unequivocally [27,41].

There is a strong belief that GC-content is the major factor in ensuring survival/selective advantages for extremophiles, in particular thermophiles, due to high thermostability of GC-pairs [40]. Assuming that this explanation is correct, one would expect (hyper) thermophiles to select Arginine over Lysine. Arginine is encoded by six codons, four of which (CGU, CGC, CGA, and CGG) are GC-rich, whereas Lysine is encoded by two codons (AAA and AAG). Moreover, Arginine has a higher charge which means it forms better salt bridges [41]. Surprisingly, this expectation is confirmed in only a very few cases, for instance in *A. pernix* and *M. kandlerii*; whereas in majority of other hyperthemophilic organisms we observe significant increase in Lysine content, which typically anticorrelates with GC-content. Furthermore, as we demonstrated here, Lysine content partially increases due to direct replacement of Arginine residues (Table 2), which points out to the obvious advantage that Lysine residues have over Arginine. One could argue, that (i) composition effect alone can account for the higher substitution rate of Arg/Lys, or that (ii) there was a particular common ancestor enriched by Lys and the specific compositional bias in contemporary proteins that we observe is due to phylogeny. But the unfolding simulations, the statistical data on rotameric states, and the genomic evidence all point to the advantage of Lysine over Arginine when thermostability is important. Furthermore we see excess of Lysine only in hyperthermophilic organisms regardless of their loci on phylogenetic tree (e.g. archaea and bacteria). Lysine still has some entropic freedom even in the folded state of a protein due to its smaller size. In comparison, Arginine, with its bulky guanidinium group, does not have the same freedom, and its possible enthalpic advantage is compensated by the drawback of packing of two closely located charges [42].

***Adaptation to high temperatures as a complex effect of different types of interactions***

We discovered here a novel mechanism of structure thermostabilization that relies on side-chain rotamer entropy [17,21]. To single out the potential effects of rotamer entropy we compared pairs of amino acid residues with similar physical and chemical properties and the same maximal number of rotameric states. The difference in the rotamer entropy of each pairs of residues must, then, be a result only of the difference in the rotameric entropy of their side-chains. Statistical data of accessible rotameric states (see Figures 2, 3 and Table 1) show substantial entropy for Lysine residues in both folded and unfolded states, whereas Arginine has a significantly decreased side-chain freedom in the folded state. Preference for the Lysine is also supported by the genomic data (see Figure 6 and Table 2) and illustrated by the computational mutation experiment (Figures 4 and 5).

In general, just a few mutations can make the difference between a mesophilic protein and its (hyper)thermophilic counterpart. Stability is reached by fine tuning sequences/structures rather than by drastic rearrangement. Moreover, in the case of hyperthermophilic proteins, practically all possible means of stabilization appear to be utilized. Any additional element of stabilization must both preserve the already-achieved level of stability and provide additional stabilization by invoking only minor modifications in sequence and structure. Arg/Lys replacement satisfies both of these conditions and, thus, exemplifies using the entropic contribution while simultaneously preserving the charged nature of the residues, which is important for other mechanisms of stabilization, such as the electrostatic [43,44].

The novel mechanism of thermal stabilization reported here is unique in a sense that it relies not only on the physical and chemical properties of a residue, but also on its dynamics in folded state effecting its entropic contribution. And because the effect is small, it can be revealed only in careful simulations and genomic comparisons. The important pedagogical point we draw from this result is that the study of protein stability on individual proteins using current state-of-the-art energy functions may result in missing subtle thermodynamic evolutionary signals that only become apparent in high-throughput analysis of proteomes and genomes.

## Materials and methods

### Statistics of rotameric states

The number of accessible rotamers in the folded (T=1) and fully unfolded (T=4, see Figure 1) states for a representative set of proteins was calculated. The temperature dependence for the number of accessible rotamers in Hydrolases H from *E. coli* (1INO) and *T. thermophilus* (2PRD) was calculated at absolute temperatures T=1, 2, 3, 3.5, and 4. Structure coordinates were recorded at every $10^5$ MC steps for a total of $10^7$ steps. The number of rotamers for every residue were determined as an average over 100 snapshots.

We used the following PDB structures to collect statistics of rotameric states (see also Table S1): 1. Hydrolases (1INO and 2PRD); 2. Rubredoxins (1RDG, 5RXN, 8RXN, and 1CAA); 3. 2Fe-2S Ferredoxin (4FXC, 1FRR, 1FRD, 1DOI, and 2CJN); 4. 4Fe-4S Ferredoxin (1FCA, 1DUR, 1IQZ, and 1VJW); 5. Chemotaxis protein (3CHY, 2CHF, and 1TMY).

Statistics of rotameric states in original and mutated structures of Hydrolases (1INO and 2PRD) and Cytochrome C (1DW0) were collected from recorded structures at every $10^4$ MC steps for a total of $10^7$ steps done for every original/mutated structure (1000 snapshots).

### High-throughput sequence analysis

We used the BLAST program[45] to create a set of pair wise alignments with significant e-value (e=0.05) using the substitution matrix BLOSUM62. We chose only sequences that had gaps of length 3 or less, and full alignment length of 45 residues or more.

### Molecular dynamic minimization

We used the CHARMM program [46] to minimize the structure upon Arg/Lys replacement. CHARMM minimization was done using the following procedure. Hydrogen coordinates were calculated by bond geometry and inserted into the starting structure; SHAKE was turned on for updating hydrogen positions. A generalized-Born solvation energy function (GBorn) and a dielectric constant with linear distance-dependence were used for dynamics. The residue of interest, and all atoms within a 5 angstrom radius of any atom in that residue, were permitted to move with CHARMM degrees of freedom to ensure that the mutated residue could repack locally. The dynamics simulation was initially constrained to the native state using a harmonic potential. The artificial harmonic constraint was reduced to zero slowly over consecutive cycles of adopted-basis Newton-Raphson (ABNR) minimization. To detect the effects of mutation, we minimized both mutated and original structures with the same protocol in order to use the latter one as a control.

**Unfolding Monte Carlo simulations of modified proteins**

Unfolding simulations were performed using an all-atom Gō model developed earlier [47]. In the Gō interaction scheme, atoms that are neighbors in the native structure are assumed to have attractive interactions. Hence the Gō model of interactions is structure-based. Every unfolding run consists of $2 \cdot 10^6$ steps in the unfolding simulations of Hydrolases (1INO and 2PRD, Figure 1) and their mutants (Figure 5a-c), and $5 \cdot 10^7$ steps in the case of Cytochrome C (1DW0, Figure 6) . The move set is one backbone move followed by one side-chain move [47].

## Acknowledgements

INB is supported by Merck Postdoctoral Fellowship for Genome-related research. This work is supported by NIH.

## Legends to tables

**Table 1.** Average number of rotamers per residue type in the folded state (absolute temperature T=1) in Hydrolases H from *E. coli* (1INO) and *T. thermophilus* (2PRD). Bold-typed are maximal numbers of rotamers in residues with the same (*e.g.* Ile/Leu or Ser/Thr) or similar (Phe/Tyr or Thr/Val) physical and chemical features. Red bold-typed are parameters for the pair Arg/Lys.

**Table 2.** Percentage of the forward/back replacements in alignments of hyperthermophilic genomes against mesophilic one (*E. coli*). RK/KR, replacement Arg-to-Lys versus Lys-to-Arg, where the first amino acid type is in mesophilic sequence and the second one is in hyperthermophilic part of the alignment. In parenthesis are ratios of forward to back substituions and p-values calculated according to $\chi^2$ criteria where there is a statistically significant bias between forward and back replacement.

**Table 1**

| **Residue name (max number of rotamers[16])** | **Average number of rotamers per residue type in 1INO (175)** | **and 2PRD (174)** |
|---|---|---|
| ARG (**81**) | **3.5** | **5.6** |
| ASN (18) | 4.6 | 2 |
| ASP (6) | 3.6 | 3.6 |
| CYS (3) | 1 | 1 |
| GLN (36) | 12 | 7.4 |
| GLU (27) | 8.5 | 8.3 |
| HIS (9) | 1.6 | 2 |
| ILE (**9**) | 1.3 | 1.5 |
| LEU (**9**) | 1.4 | 1.4 |
| LYS (**81**) | **20.1** | **17.2** |
| MET (27) | 1.7 | 1.5 |
| PHE (**6**) | 1.3 | 1 |
| PRO (2) | 1 | 1 |
| SER (**3**) | 2.3 | 1.8 |
| THR (**3**) | 1 | 1 |
| TRP (9) | 1 | 1 |
| TYR (**6**) | 1.6 | 1.6 |
| VAL (**3**) | 1 | 1.1 |

**Table 2**

| HT genome | RK/KR | LI/IL | TS/ST |
|---|---|---|---|
| *A. aeolicus* (B) | 20.0/8.1 (2.47, $p<10^{-3}$) | 14.2/19.3 (0.74) | 7.5/6.8 (1.11) |
| *M. jannaschii* (A) | 22.4/6.0 (3.73, $p<10^{-3}$) | 20/16.7 (1.20) | 7.0/6.5 (1.08) |
| *N. equitans* (A) | 23.7/6.0 (3.95, $p<10^{-3}$) | 19.5/19 (1.03) | 6.8/6.8 (1) |
| *P. furiosis* (A) | 16.5/9.9 (1.67, $p<5\cdot10^{-2}$) | 16.3/18.2 (0.90) | 7.8/7.5 (1.04) |
| *S. tokodaii* (A) | 18.2/7.4 (2.46, $p<10^{-3}$) | 18.5/17.8 (1.04) | 9.8/7.3 (1.34) |
| *A. pernix* (B) | **8.1/15.6 (0.52, $p<10^{-2}$)** | **10.7/20.9 (0.52, $p<10^{-2}$)** | 9.8/7.2 (1.36) |

## Legends to Figures

**Figure 1.** The temperature dependence of the energy of unfolding for Hydrolases, from *E.coli* (1INO, black rhombuses) and *T. thermophilus* (2PRD, red squares). Every simulation of unfolding started from the native structure and lasted for $2 \cdot 10^6$ MC steps; absolute temperature increment is 0.2 and 0.1 in the vicinity of transition temperature.

**Figure 2.** The temperature dependence of the natural logarithm of number of rotamers. **(a)** Arg (black rhombuses) versus Lysine (red squares) rotamers of Hydrolase H from *E.coli* (1INO); **(b)** Arg (black rhombuses) versus Lysine (red squares) rotamers of Hydrolase H from *T. thermophilus* (2PRD); **(c)** Leu (dark blue rhombuses) versus Ile (light blue squares) rotamers of Hydrolase H from *E.coli* (1INO); **(d)** Leu (dark blue rhombuses) versus Ile (light blue squares) rotamers of Hydrolase H from *T. thermophilus* (2PRD); **(e)** Thr (orange rhombuses) versus Ser (yellow squares) rotamers of Hydrolase H from *E.coli* (1INO); **(f)** Thr (orange rhombuses) versus Ser (yellow squares) rotamers of Hydrolase H from *T. thermophilus* (2PRD); **(g)** Thr (orange rhombuses) versus Val (green-blue squares) rotamers of Hydrolase H from *E.coli* (1INO); **(h)** Thr (orange rhombuses) versus Val (green-blue squares) rotamers of Hydrolase H from *T. thermophilus* (2PRD).

**Figure 3.** Distribution of the ratios of the number of rotamers in unfolded and folded states in a representative set of proteins. Completely unfolded state is achived at absolute temperature T=4, folded state at T=1. **(a)** Lys versus Arg; **(b)** Ile versus Leu; **(c)** Phe versus Tyr; **(d)** Val versus Thr. Upper histogram in each panel corresponds to T=4, lower histogram corresponds to T=1

**Figure 4.** The temperature dependence of the energy of unfolding for mutated (red squares) versus original Hydrolases H: **(a)** R171K mutant and wild-type of Hydrolase H from *T. thermophilus* (2PRD); **(b)** R171K mutant and wild-type of Hydrolase H from *E.coli* (1INO); **(c)** R43,86,88,171K mutant and wild-type of Hydrolase H from *E.coli* (1INO).

**Figure 5.** The temperature dependence of the energy of unfolding for R24,26,53,58,74,80,87,95K mutant compared to the original structure of Cytochrome C from *R. sphaeroides*.

**Figure 6.** Histograms of the content of charged amino acid residues in hyperthermophilic genomes compared to mesophilic genomes. Top histogram shows percentage of each residue in mesophilic genomes; bottom histogram, in hyperthermophilic genomes. A total of 12 hyperthermophilic and 38 mesophilic genomes were analyzed (for the complete list see Tables S1 and S2 in Supplementary Materials). **(a)** Arg**; (b)** Lys; **(c)** Asp; **(d)** Glu.

**Figure 1**

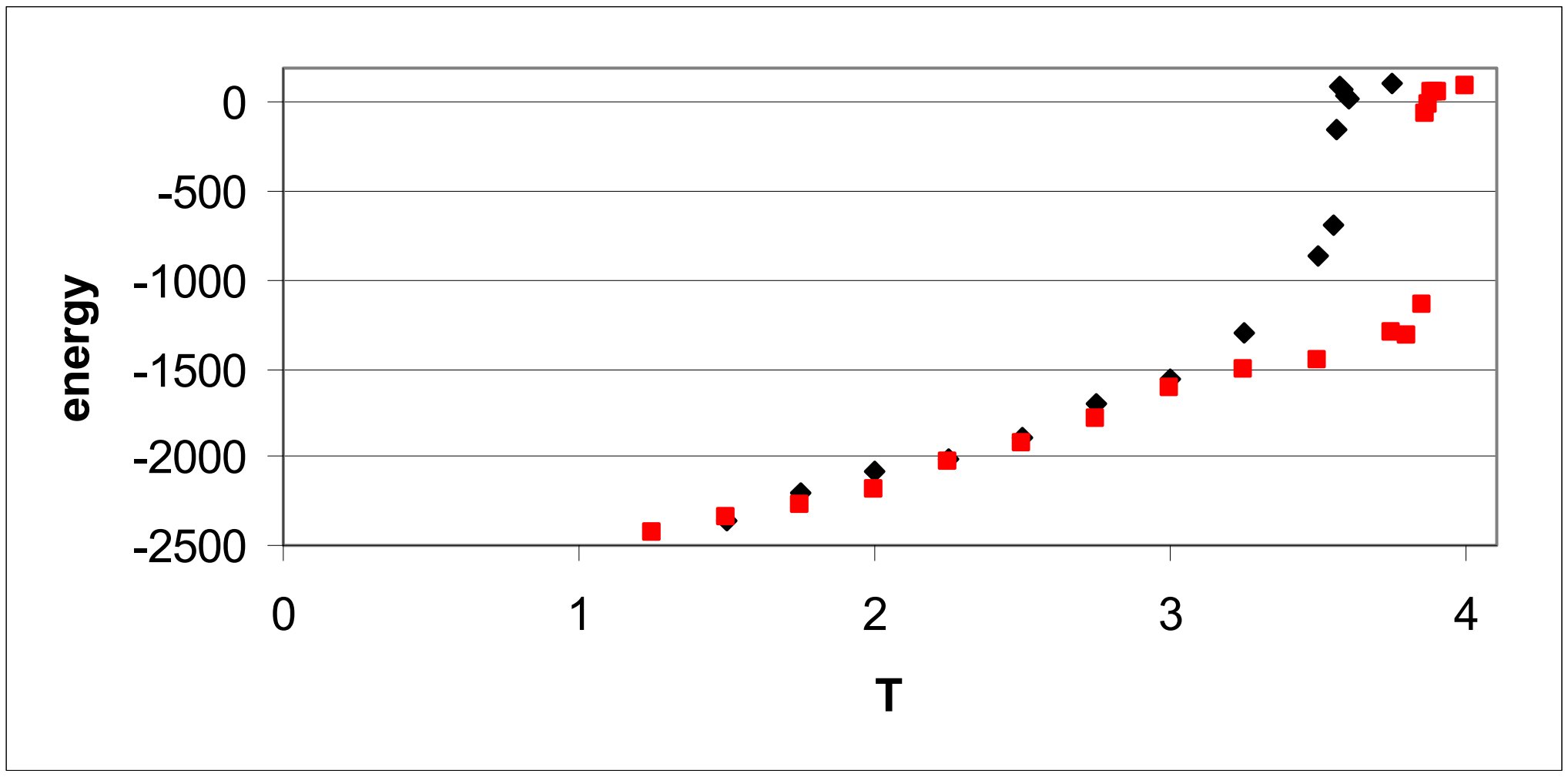

# Figure 2

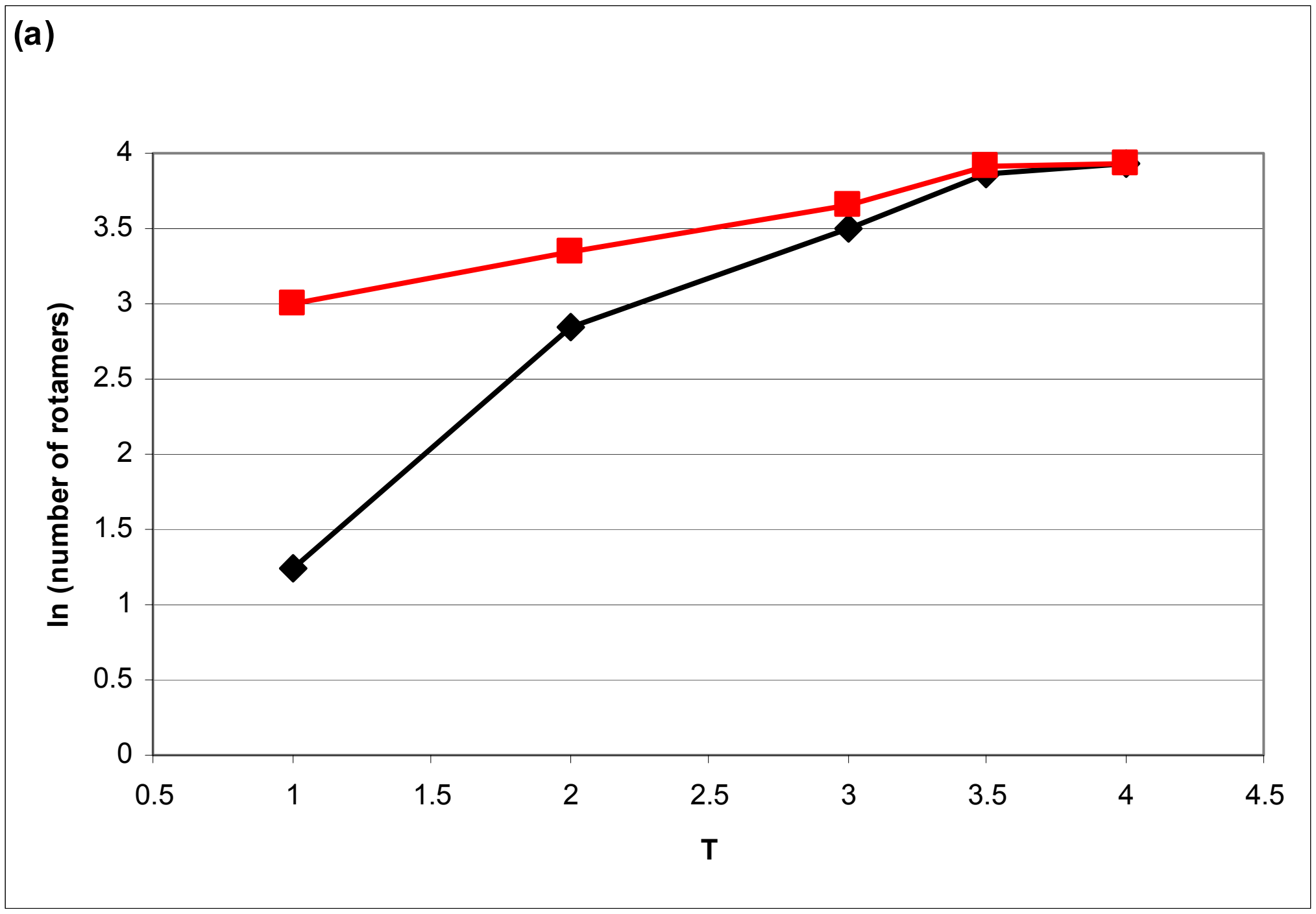

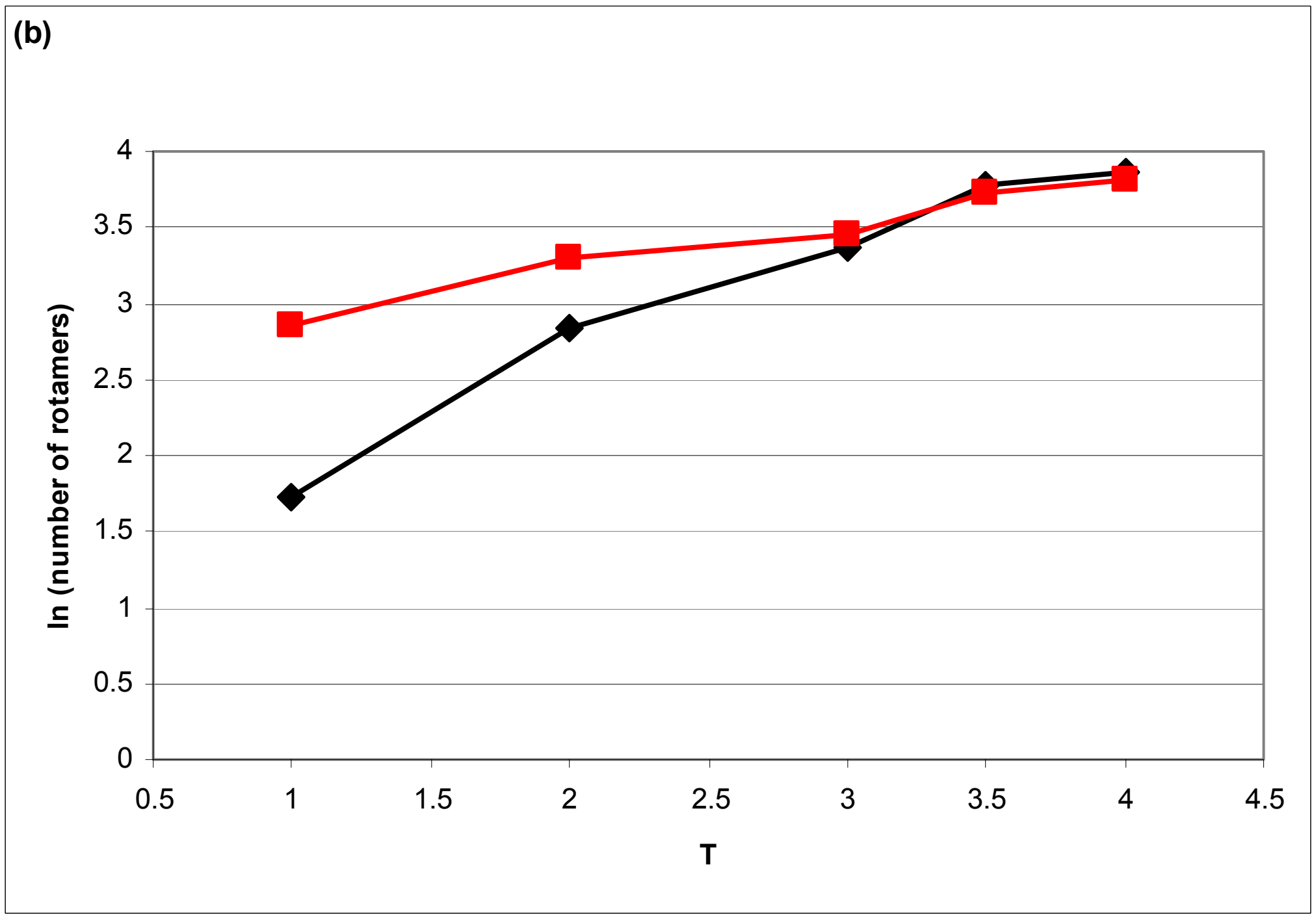

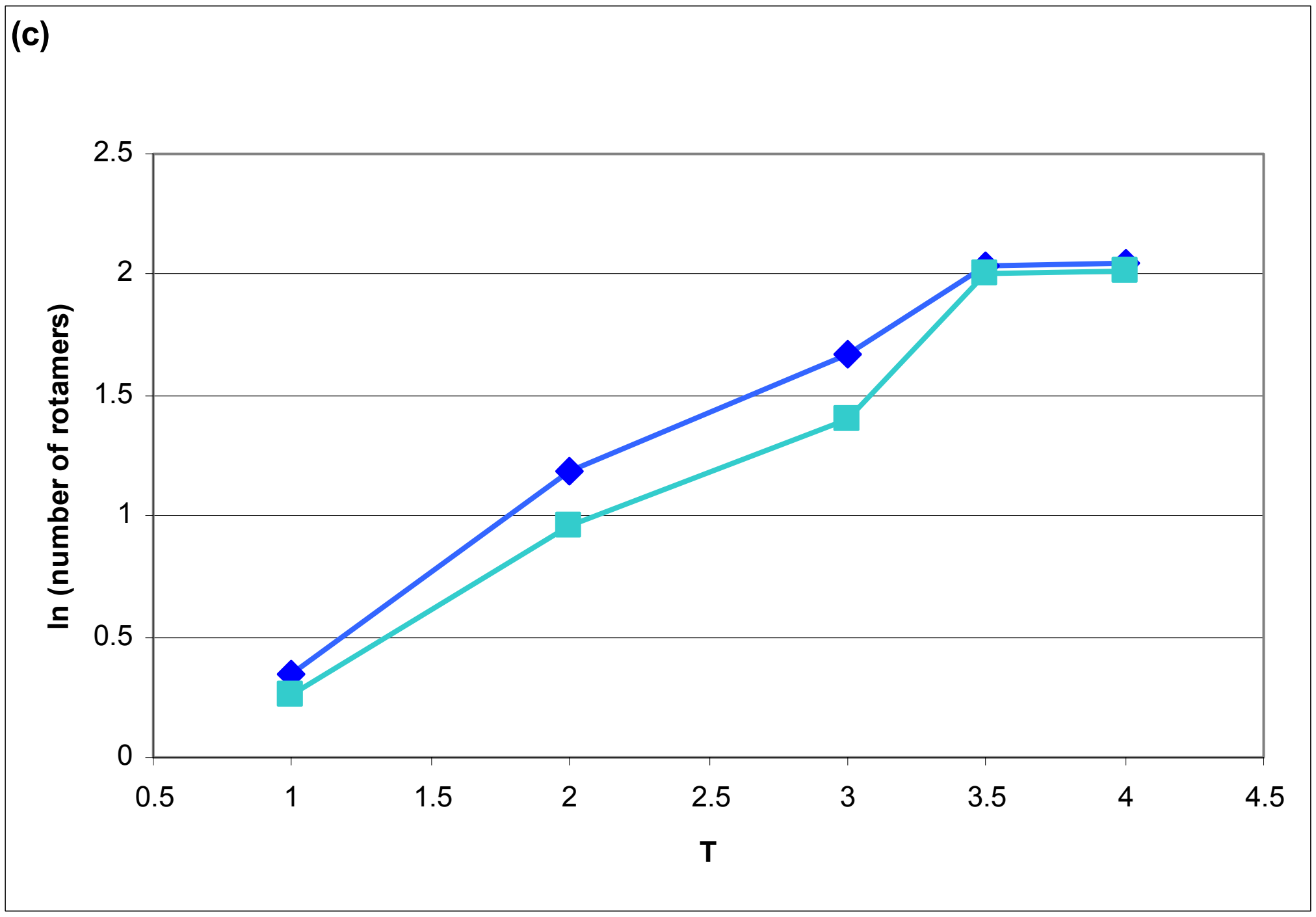

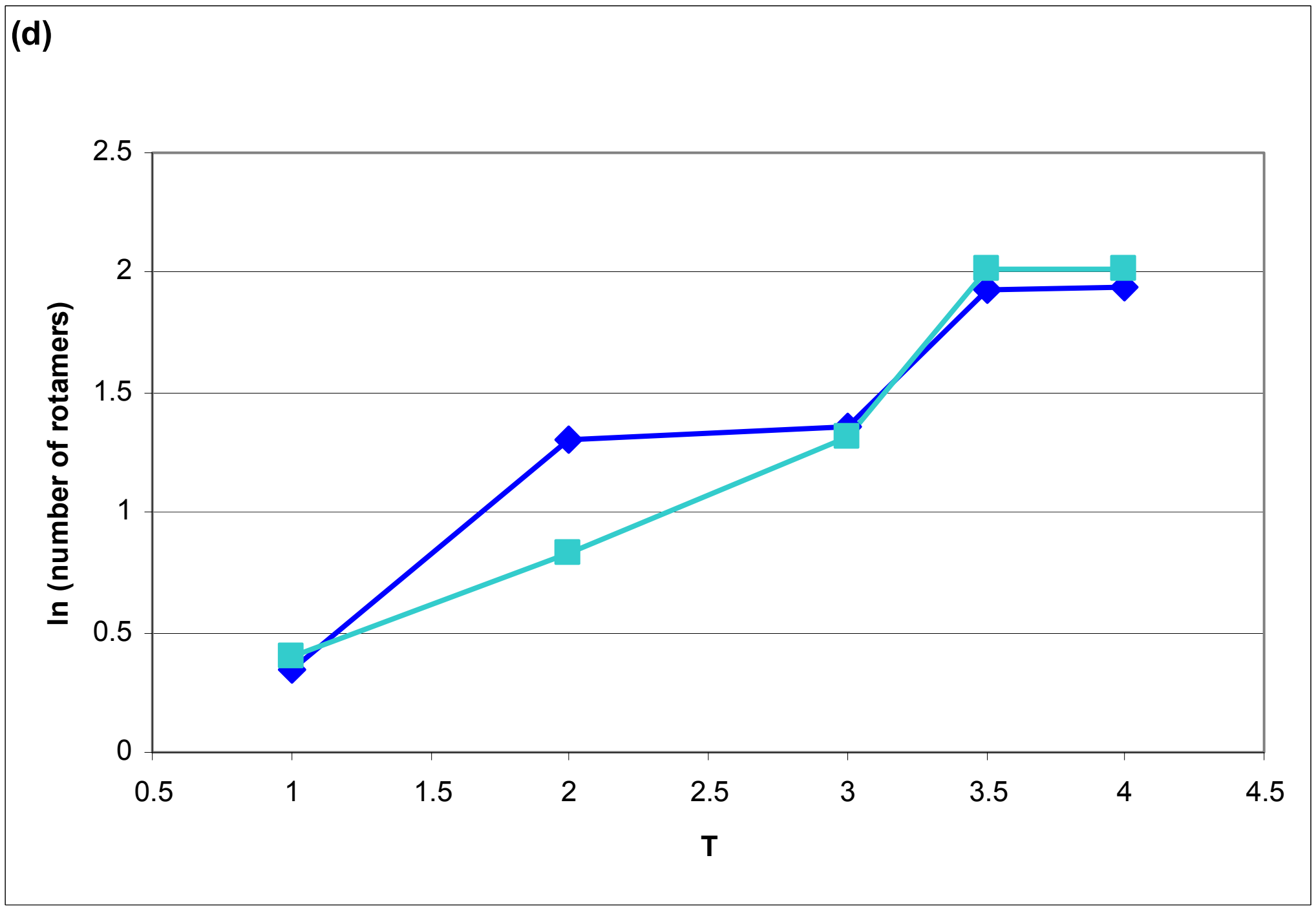

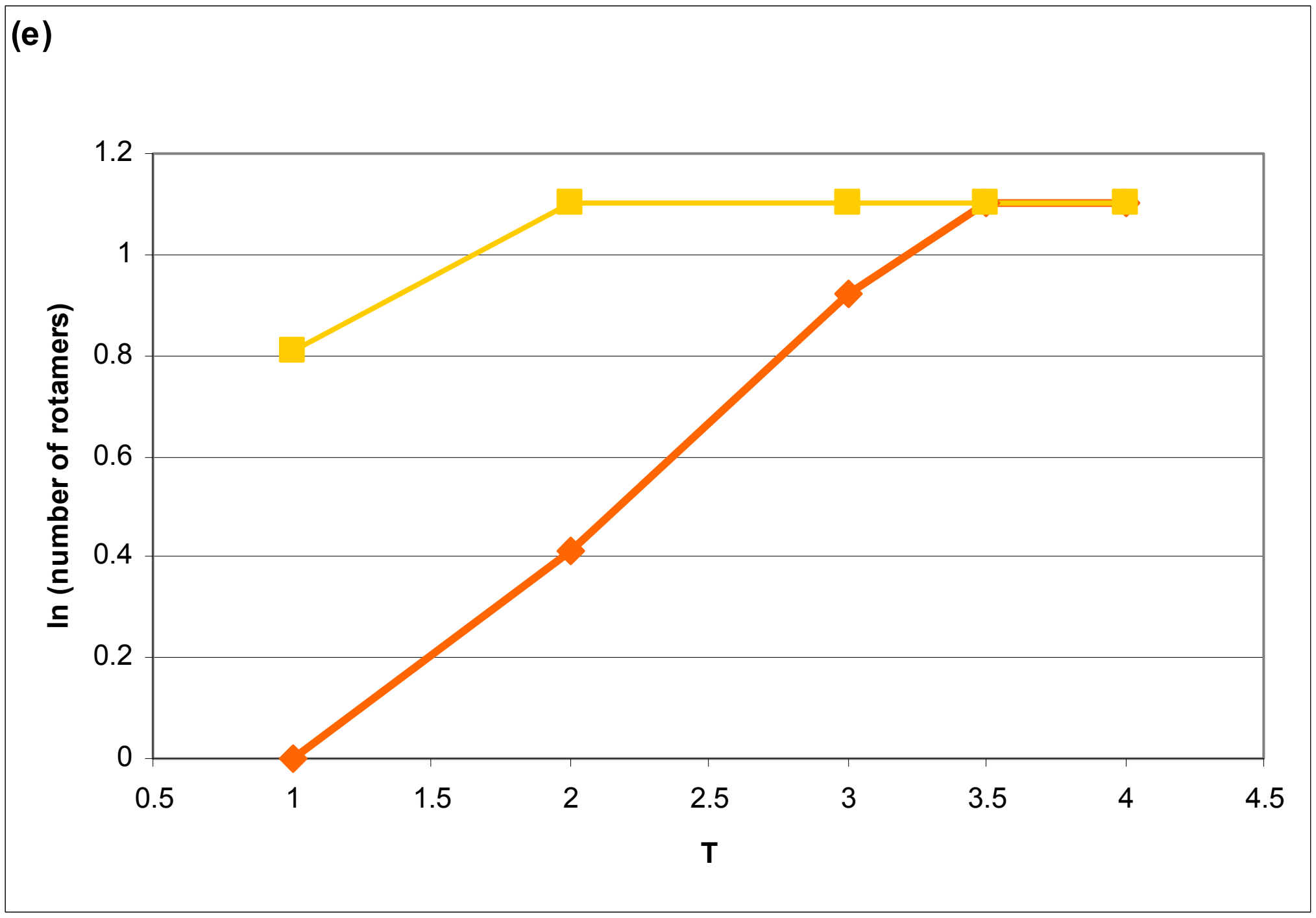

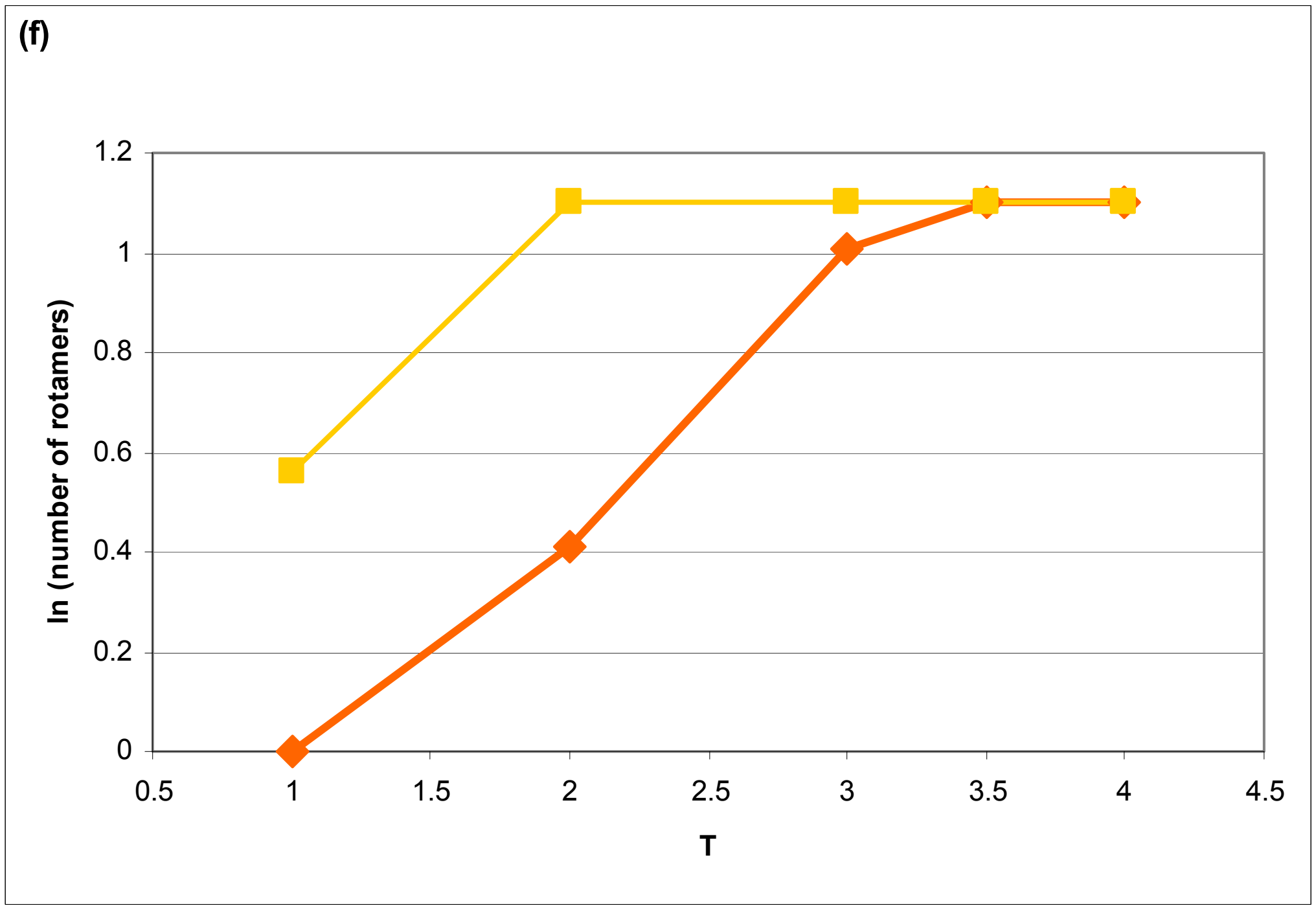

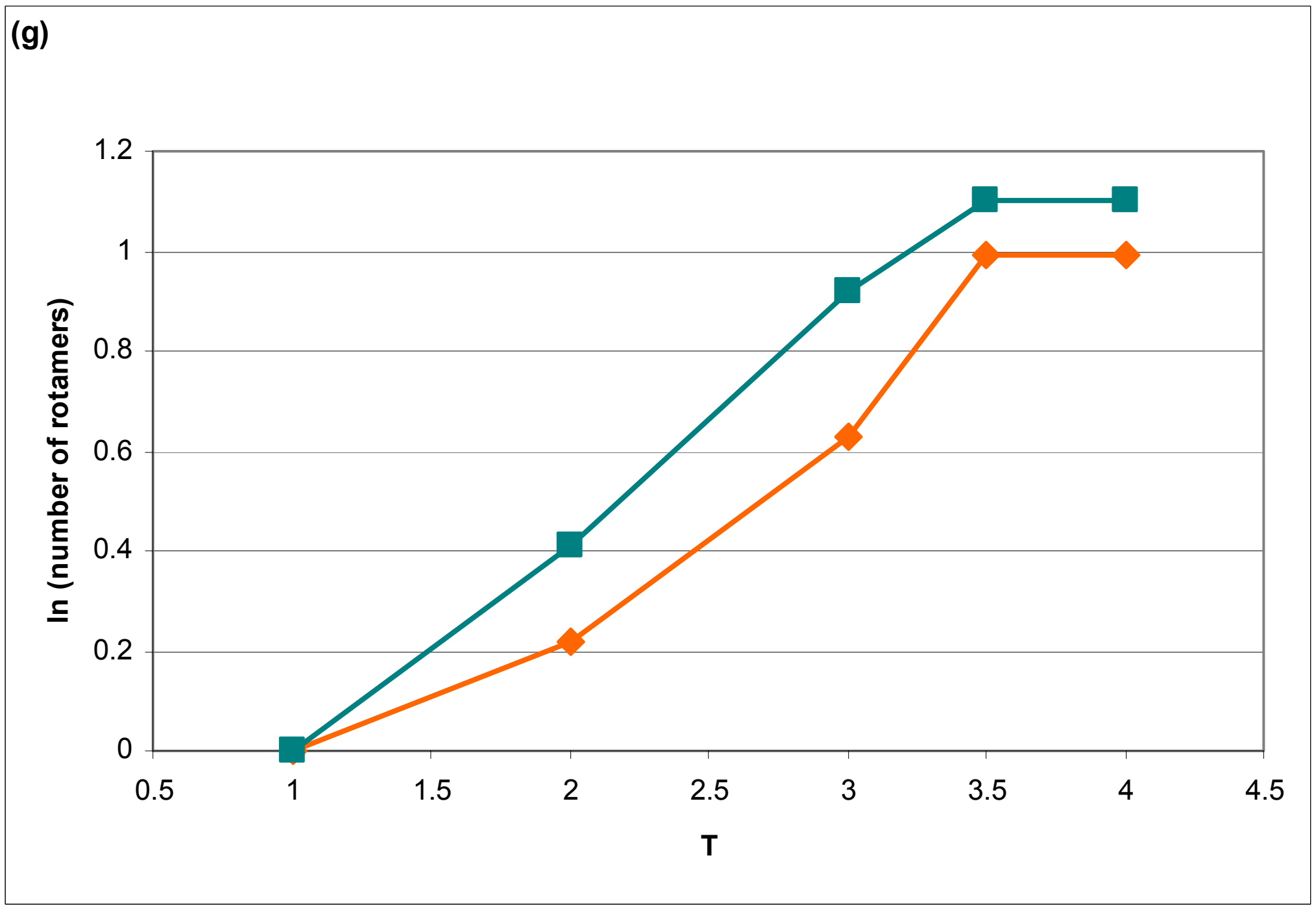

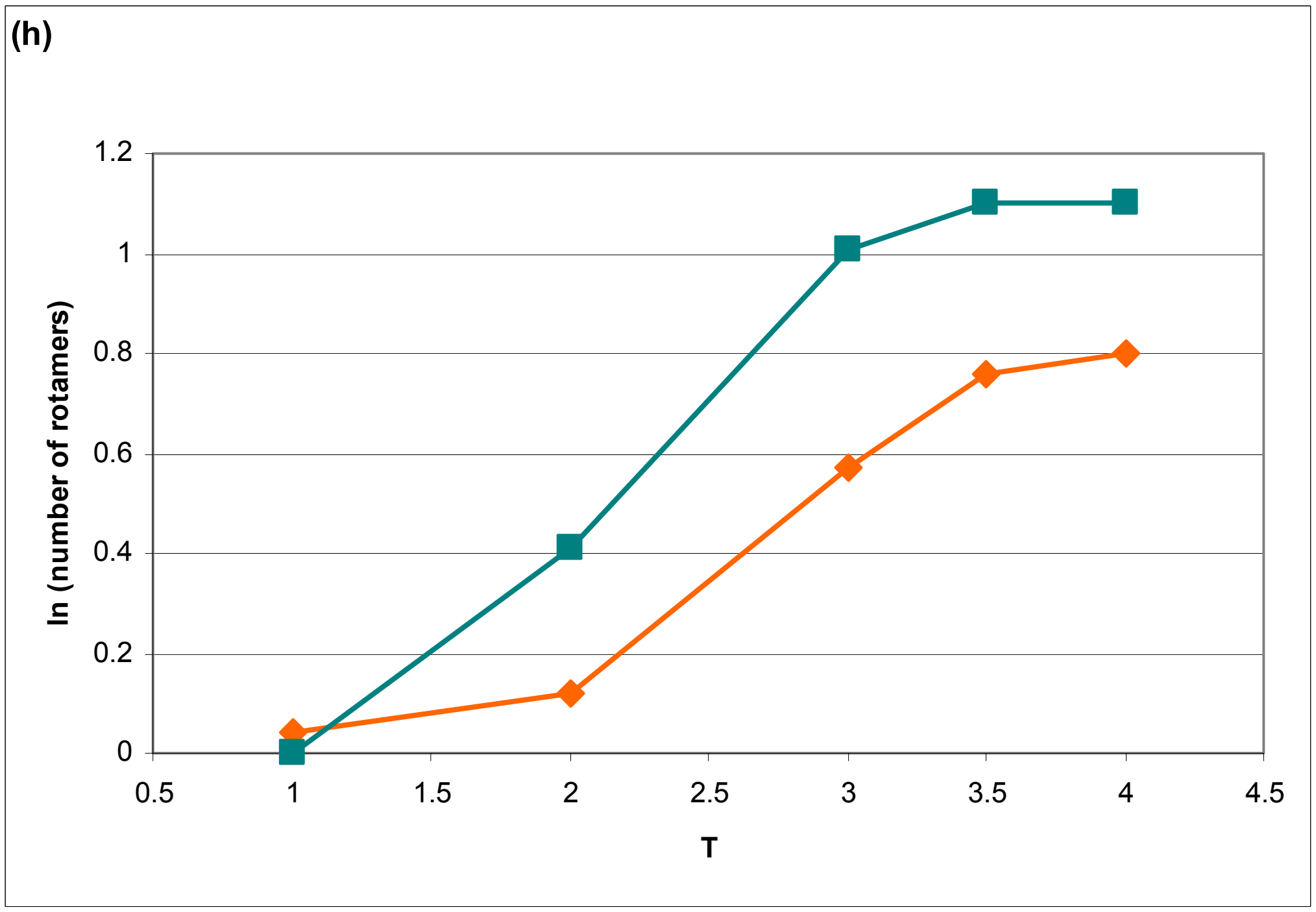

**Figure 3**

**(a)**

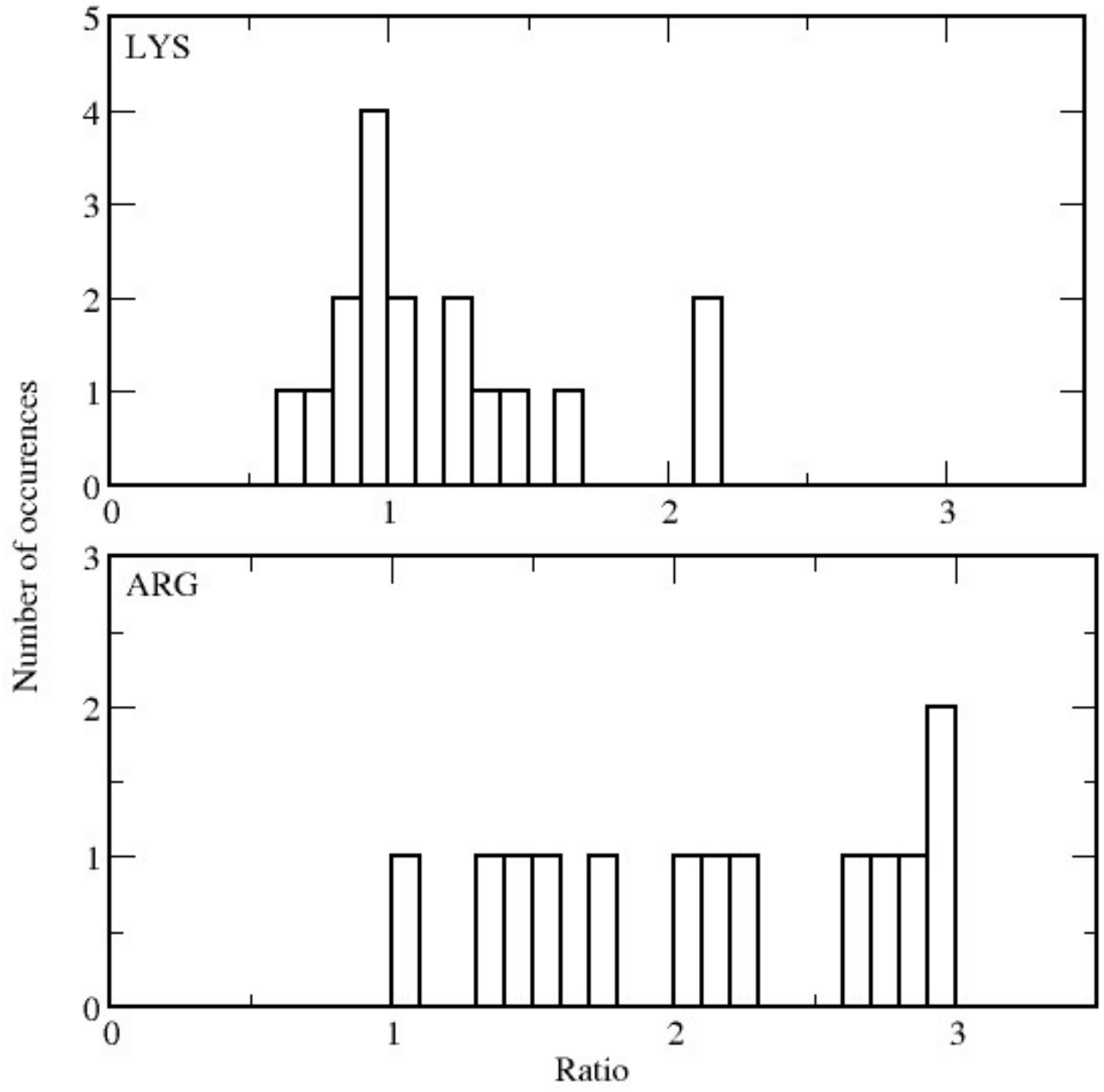

**(b)**

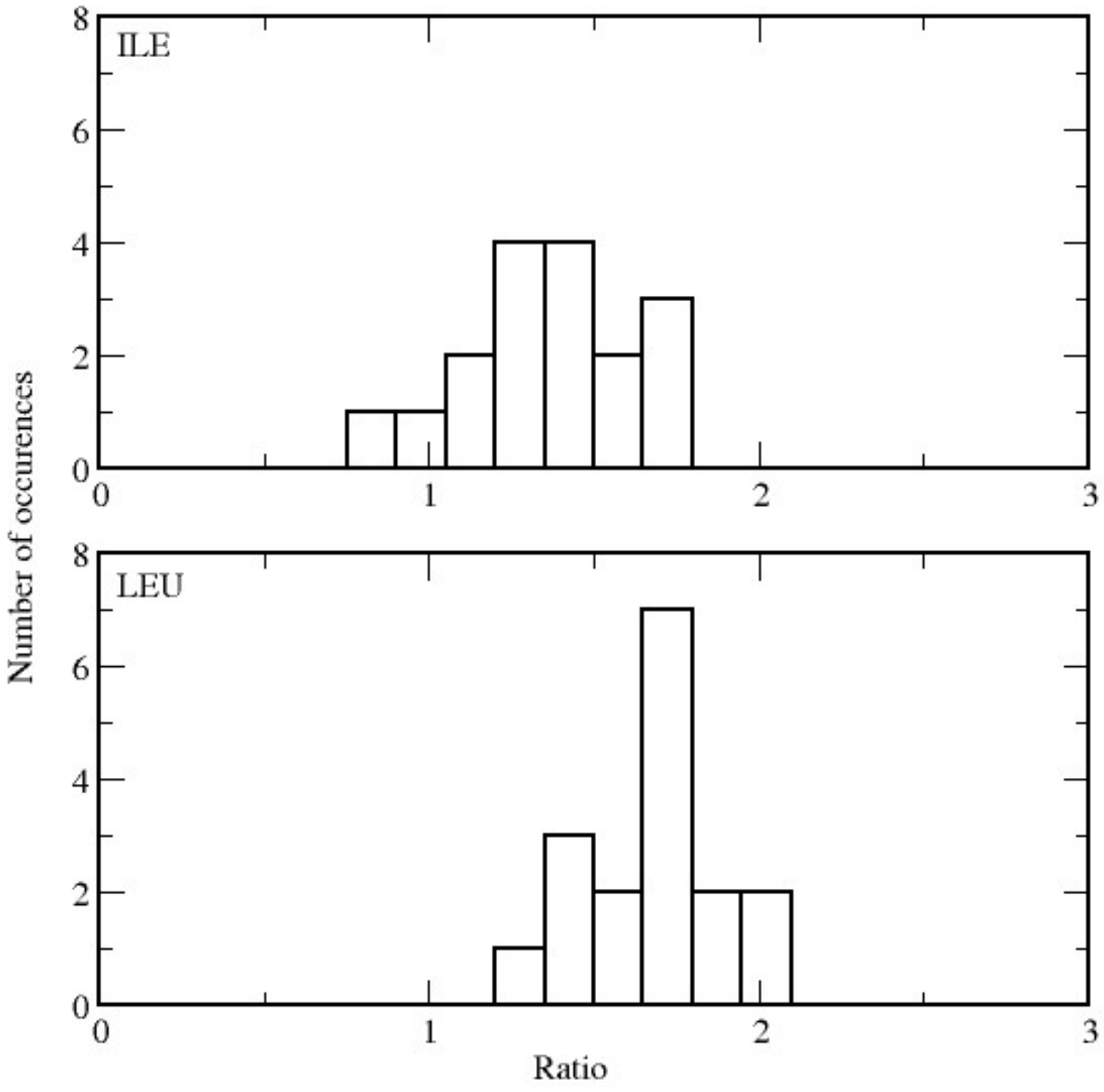

**(c)**

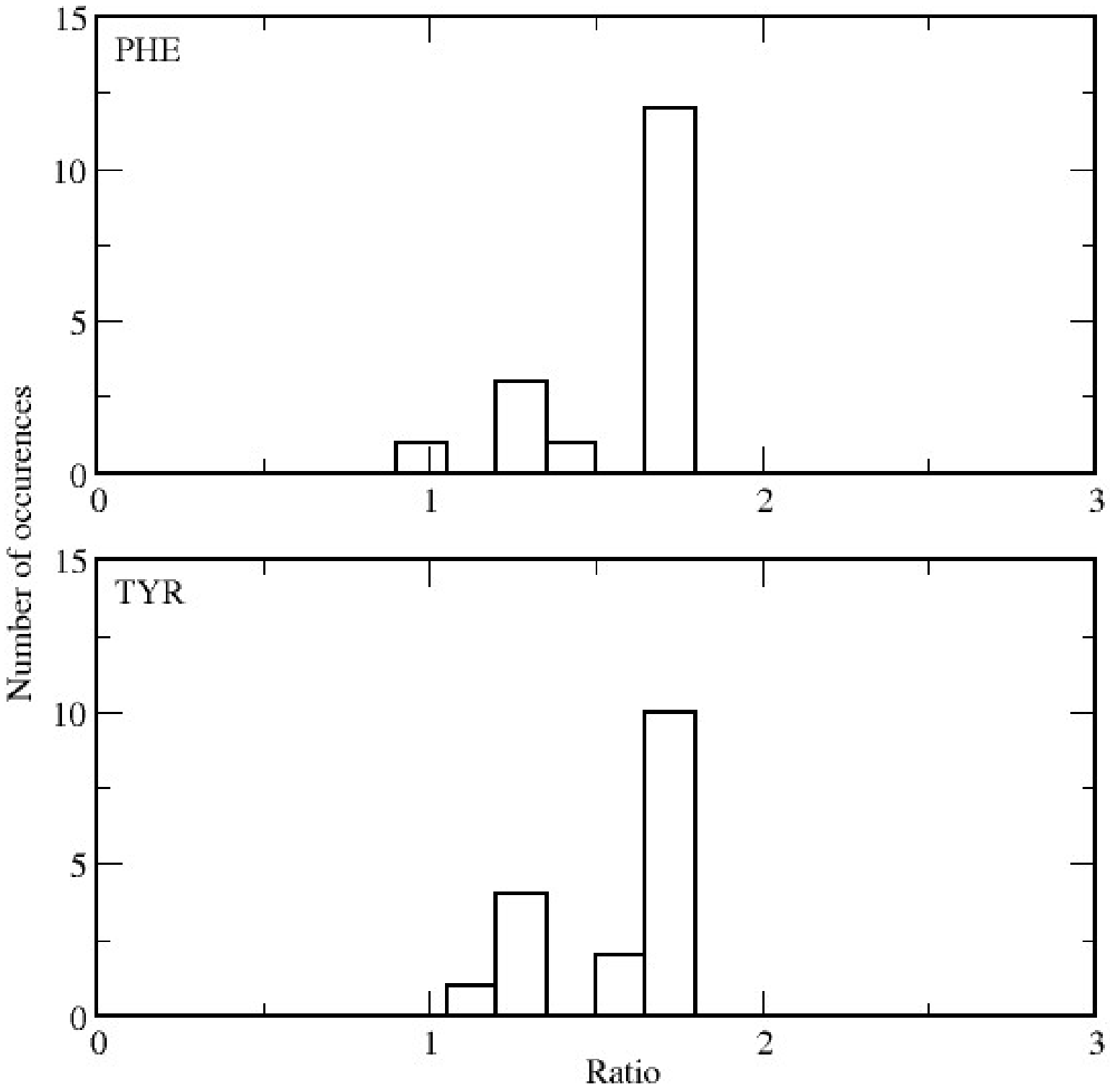

PHE
TYR
Number of occurences
Ratio

**(d)**

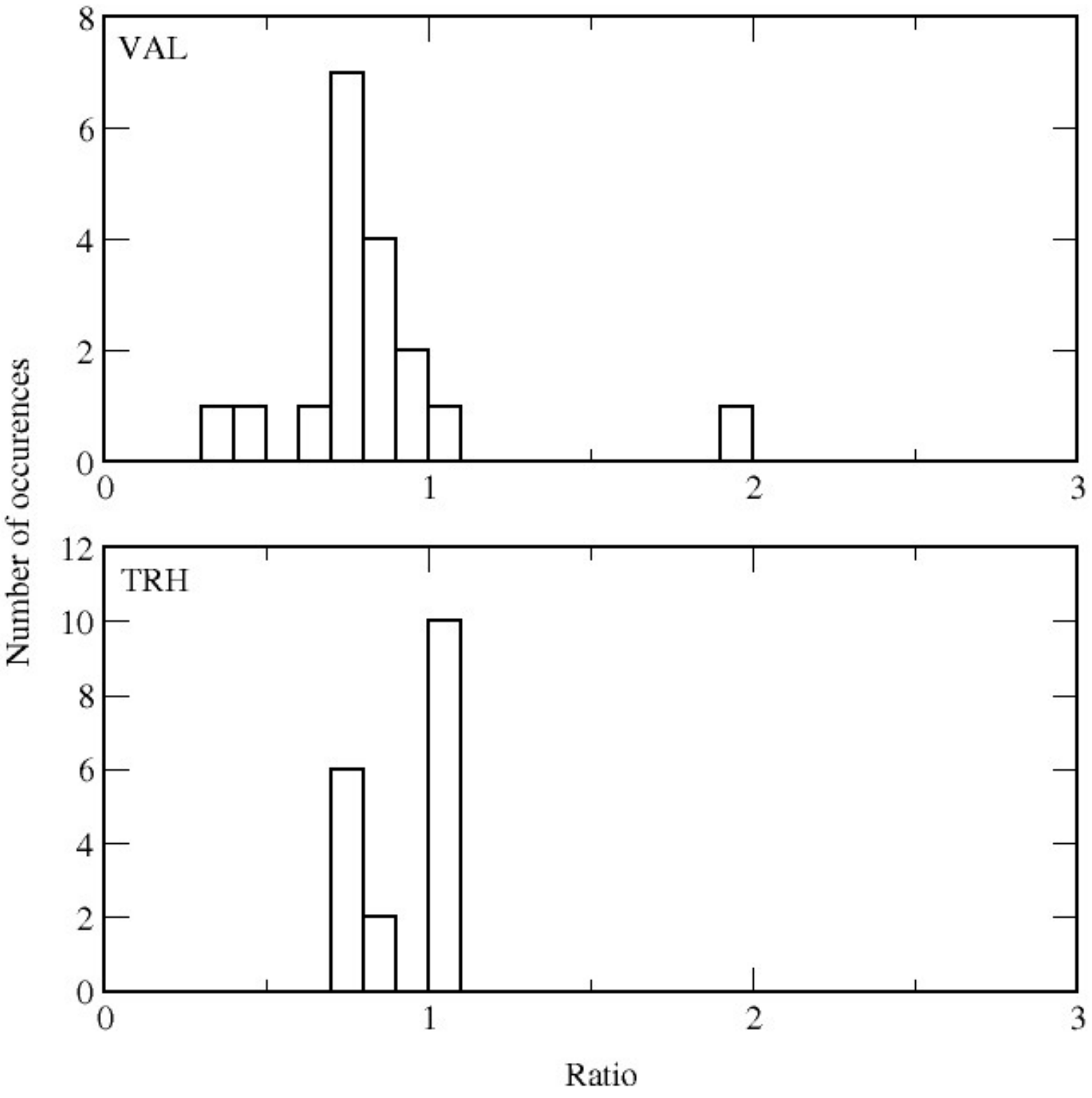

VAL
TRH
Number of occurences
Ratio

**Figure 4**

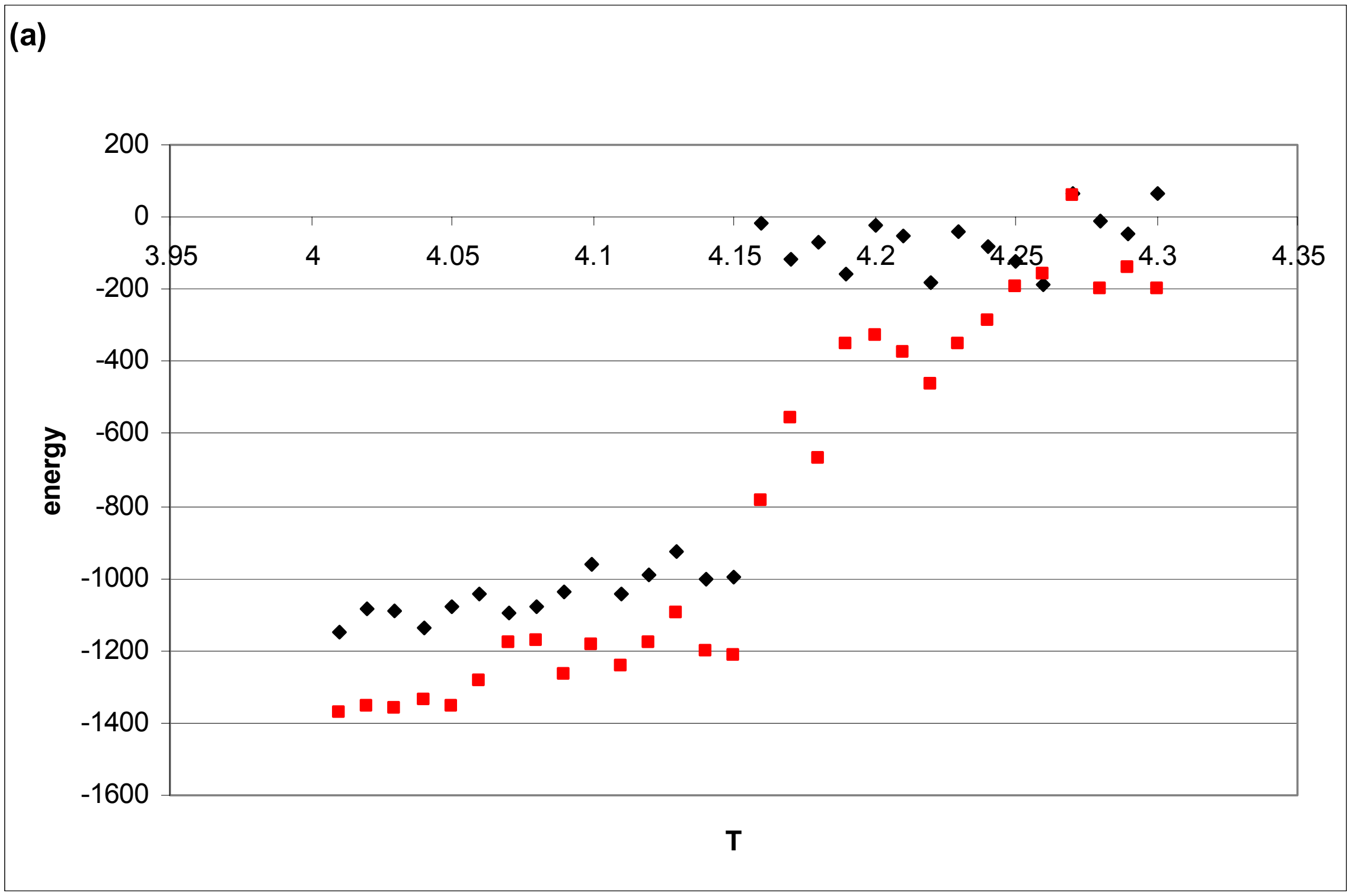

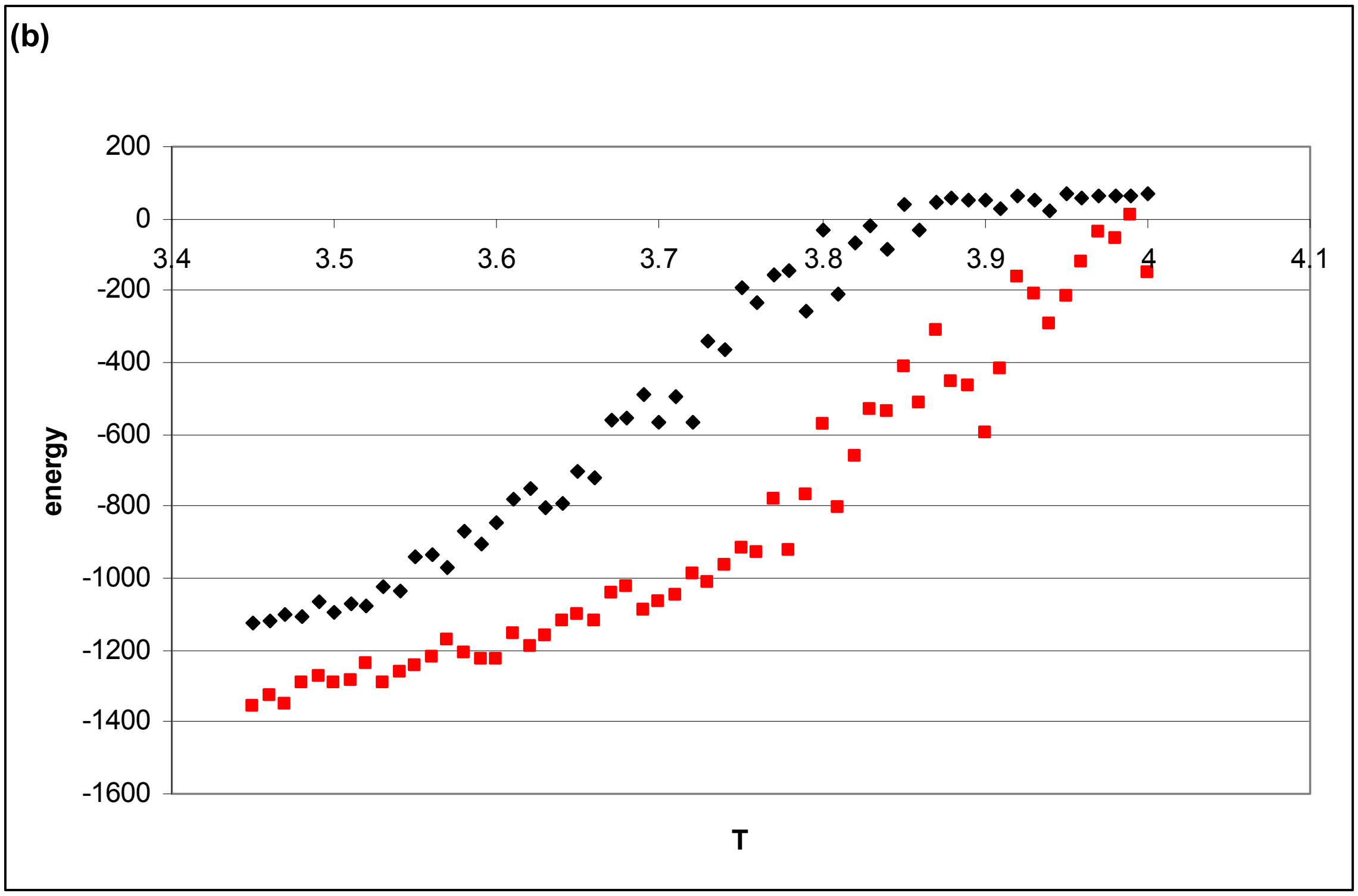

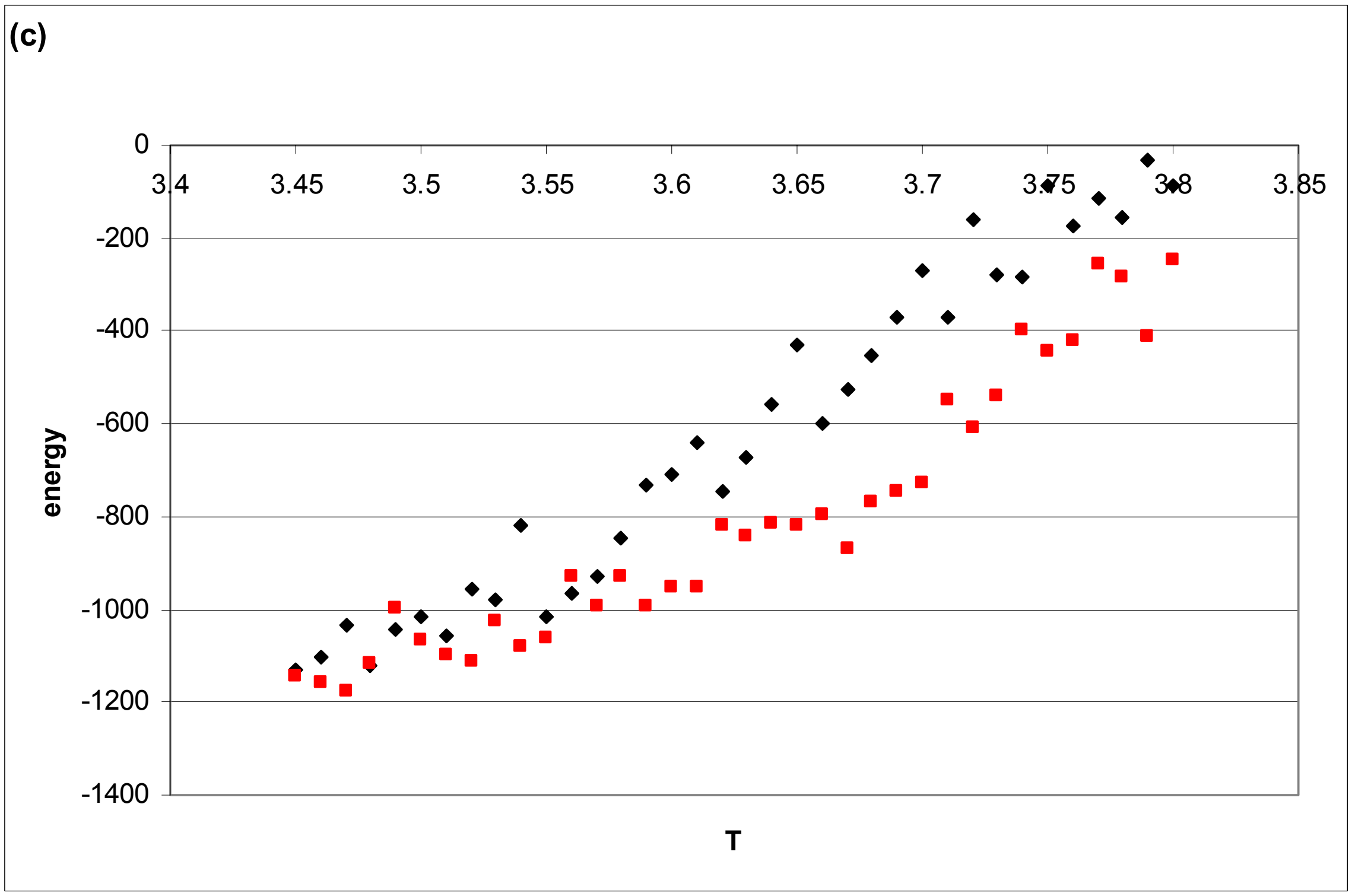

# Figure 5

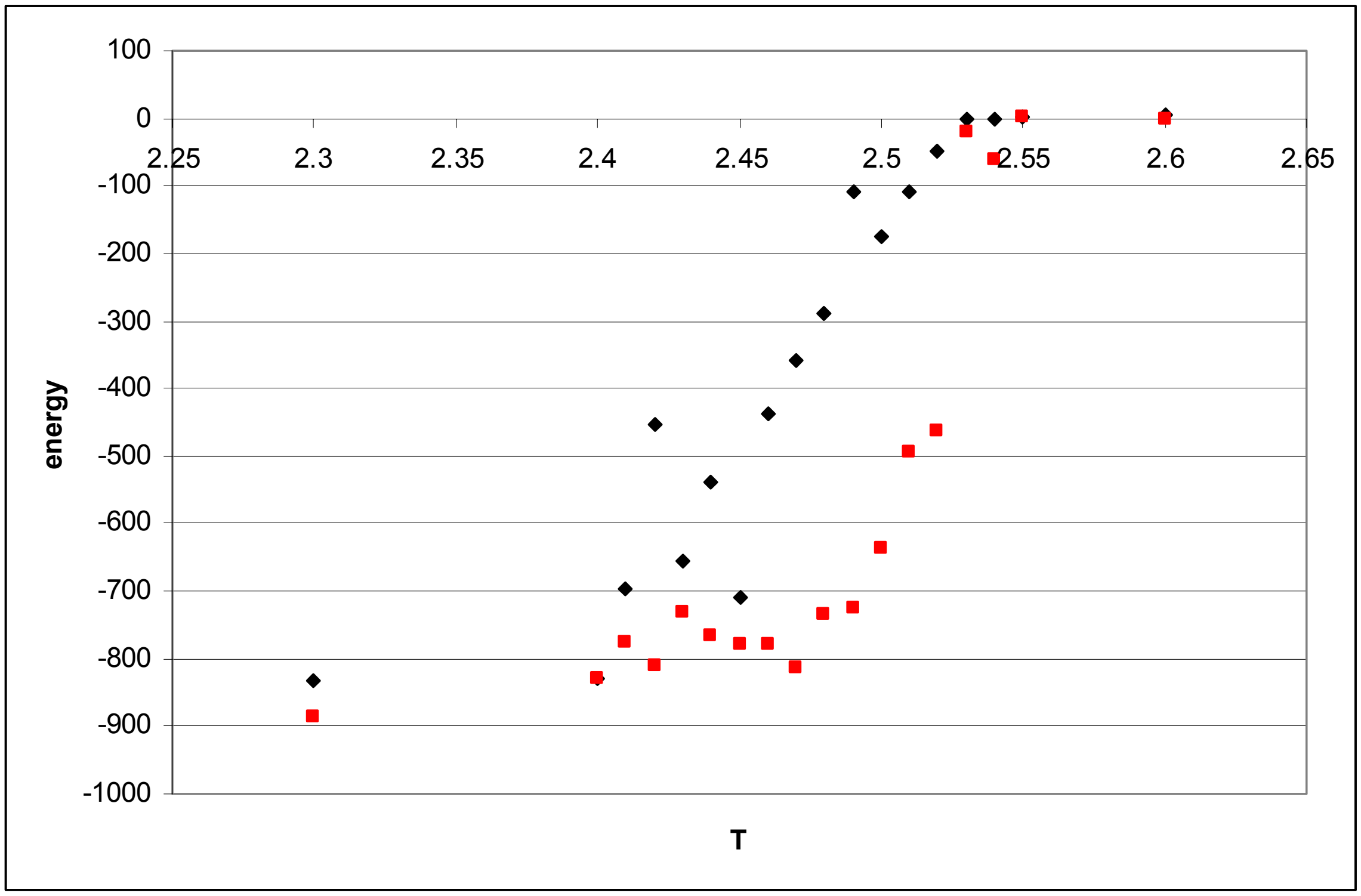

**Figure 6**

**(a)**

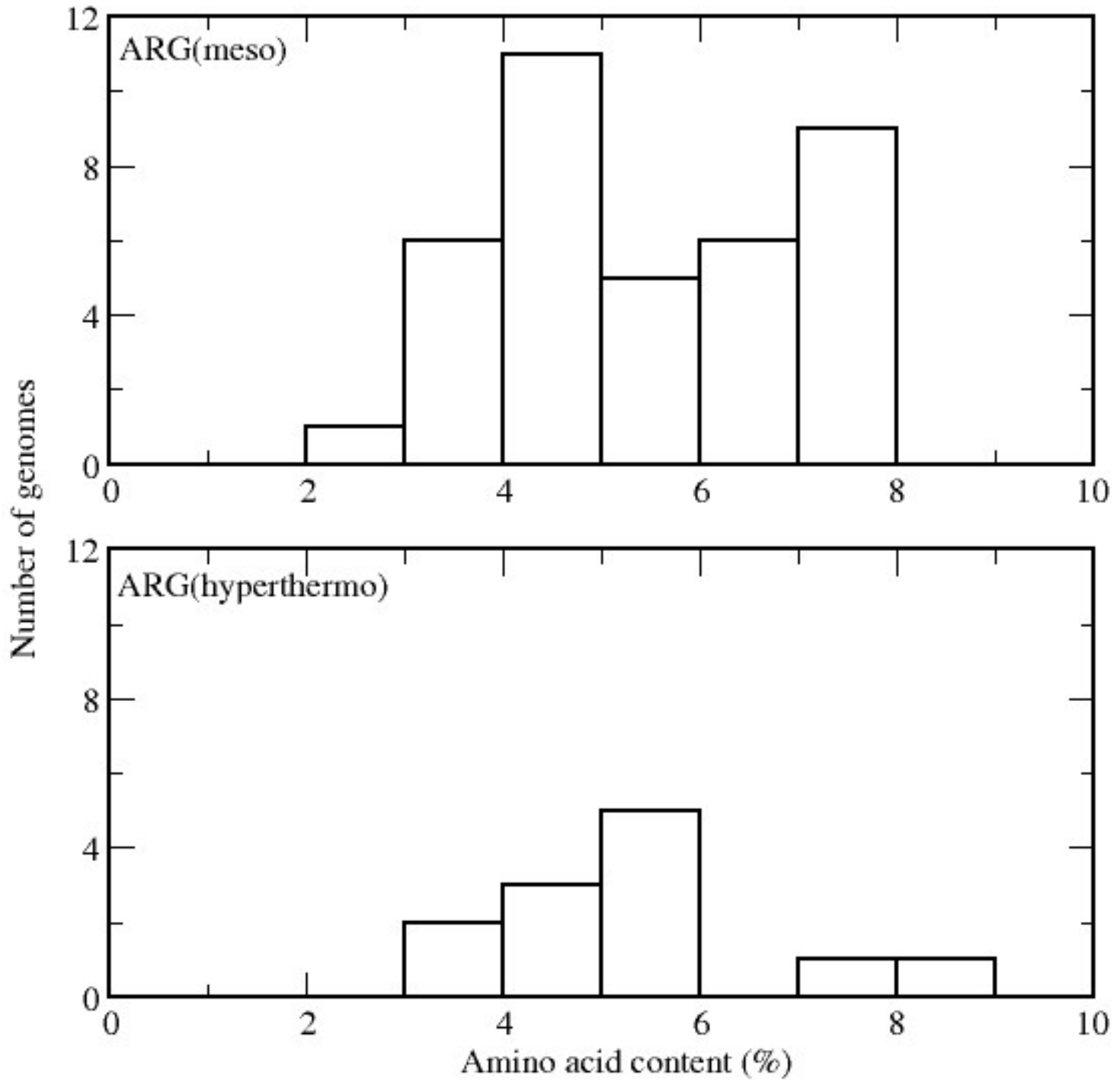

**(b)**

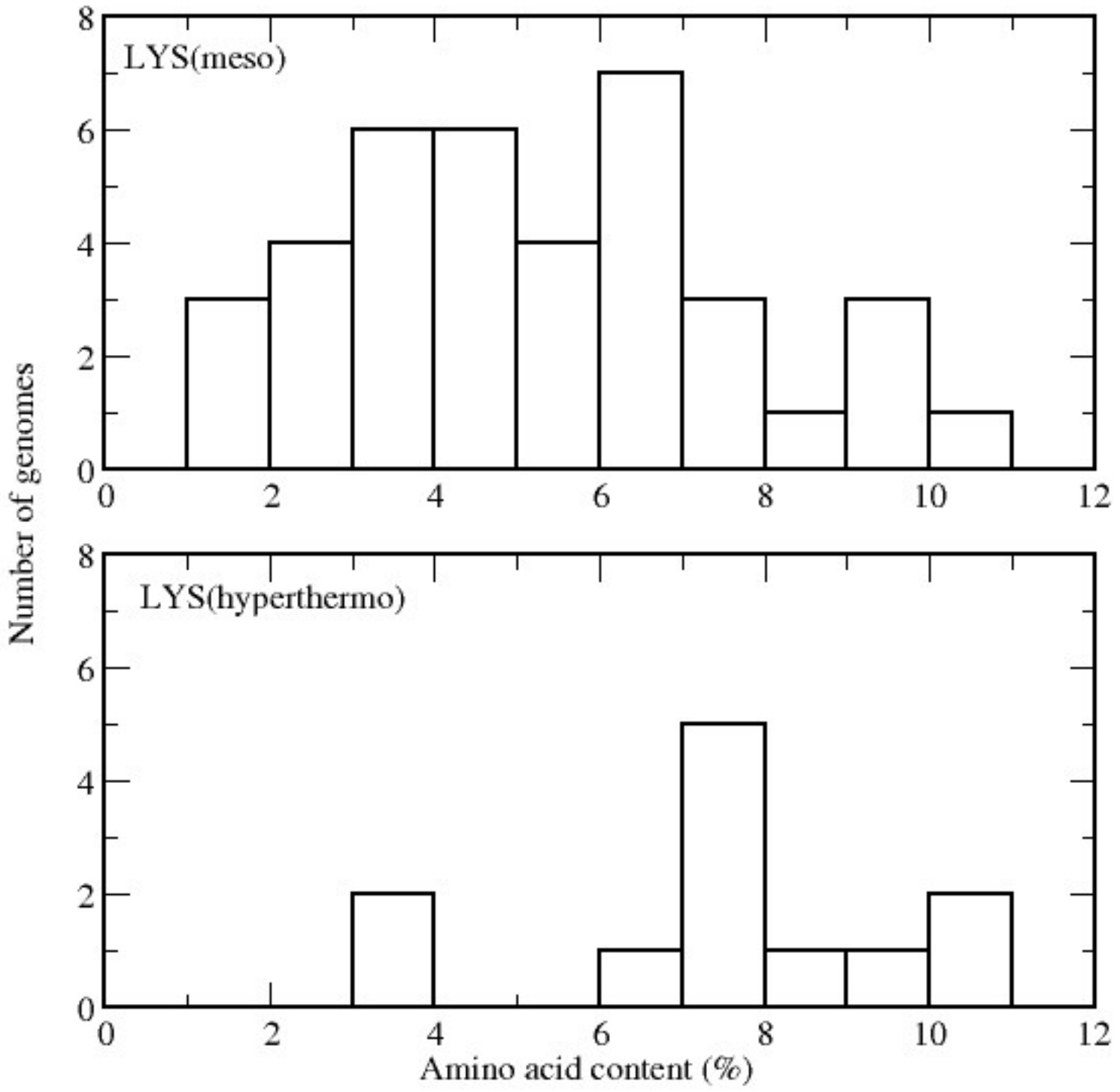

**(c)**

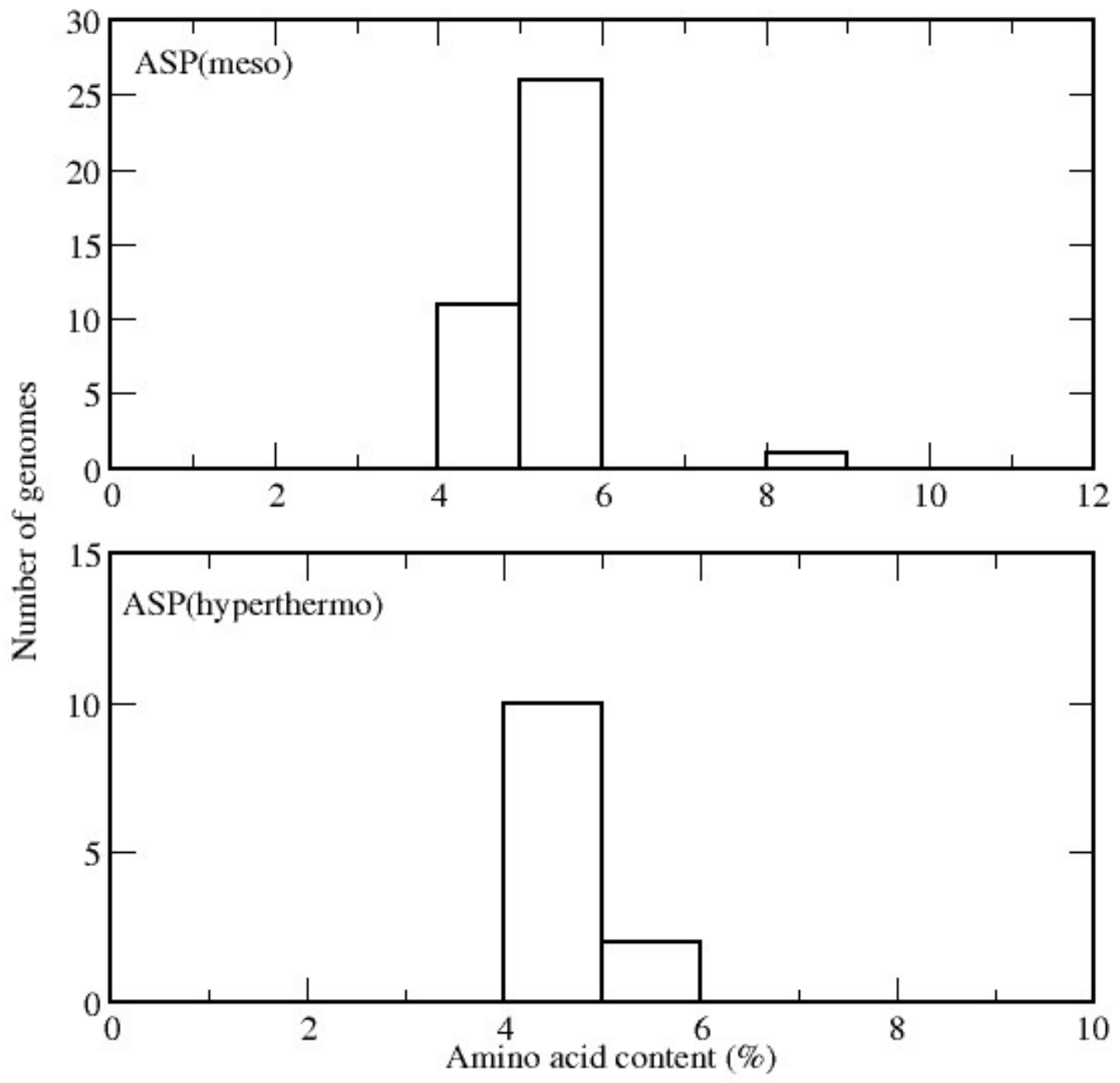

ASP(meso)
ASP(hyperthermo)
Number of genomes
Amino acid content (%)

**(d)**

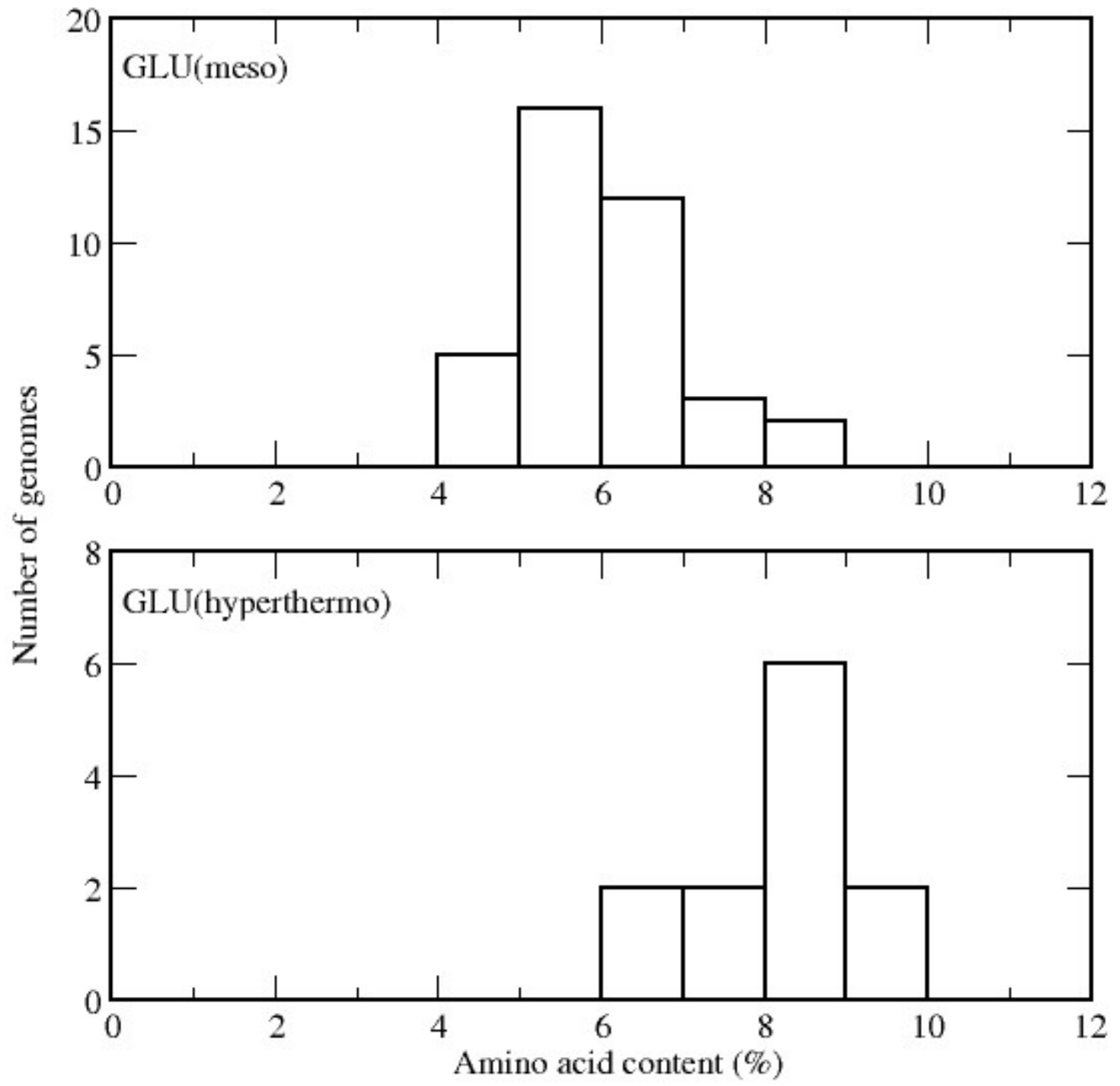

GLU(meso)
GLU(hyperthermo)
Number of genomes
Amino acid content (%)

## Supplementary materials

## Legends to tables

**Table S1.** Set of proteins used in collecting comparative rotamer statistics.

**Table S2.** List of mesophilic genomes. Total of 38 genomes. Columns are as follows: first, genome accession number in NCBI database of complete genomic sequences; second, name of the organism; third, Life Kingdom (A, archaea; B, bacteria); fourth, size of the proteome in number of protein coding sequences.

**Table S3.** List of hyperthermophilic genomes. Total of 12 genomes. Columns are as in Table S2.

**Table S4.** Expected (on the basis of the occurrence in *E. coli,* Column 4) and observed (Column 5) frequencies of charged amino acid residues in *P. furiosis*. **Diff in σ**, difference in number of standard deviation between respective expected and observed values. The null model used to calculate p-values represents random uncorrelated distribution of charged aminoacids over proteomes resulting in binomial distribution for the content of each type of aminoacids, from which p-values were calculated.

## Legends to Figures

**Figure S1.** The temperature dependence of the natural logarithm of the number of rotamers averaged over respective values in Hydrolases H from *E.coli* (1INO) and *T. thermophilus* (2PRD). **(a)** Arginine (black rhombuses) versus Lys (red squares) rotamers; **(b)** Leu (dark blue rhombuses) versus Ile (light blue squares); **(c)** Thr (orange rhombuses) versus Ser (yellow squares); **(d)** Thr (orange rhombuses) versus Val (green-blue squares); **(e)** Val (green-blue rhombuses) versus Ser (yellow squares); **(f)** Phe (green-blue rhombuses) versus Tyr (orange squares);

**Figure S2.** Histograms of the content of polar amino acid residues in hyperthermophilic genomes compared to mesophilic one. Top histogram shows percentage of respective residue in mesophilic genomes, bottom one, in hyperthermophilic ones. Total of 12 hyperthermophilic and 38 mesophilic genomes were analyzed (for the complete list see Tables S1 and S2 in Supplementary materials). **(a)** Asn**; (b)** Gln; **(c)** His; **(d)** Ser; **(e)** Thr; **(f)** Tyr.

**Table S1**

| Protein (number of residues) | Source |
|---|---|
| **Hydrolase** | |
| 1INO (175) | *E. coli* |
| 2PRD (174) | *T. thermophilus* |
| **Rubredoxin** | |
| 1RDG(52) | *D. gigas* |
| 5RXN (54) | *C. pasteuranium* |
| 8RXN (55) | *D. vulgaris* |
| 1CAA (53) | *P. furiosus* |
| **Ferredoxin (2FE-2S)** | |
| 4XFC (98) | *S. platensis* |
| 1FRR (95) | *E. arvense* |
| 1FRD (98) | *Anabaena PCC7120* |
| 1DOI (128) | *H. marismotui* |
| 2CJN (97) | *S. elongates* |
| **Ferredoxin (4FE-4S)** | |
| 1FCA (55) | *C. acidiurici* |
| 1DUR (55) | *P. asaccharolyticus* |
| 1IQZ (81) | *B. thermoproteolyticus* |
| 1VJW (59) | *T. maritima* |
| **Chemotaxis protein** | |
| 3CHY (128) | *E. coli* |
| 2CHF (128) | *S. typhimurium* |
| 1TMY (118) | *T. maritima* |

## Table S2

| Accession number | Name | Superkingdom | Size |
|---|---|---|---|
| NC_000964 | *Bacillus subtilis* | **B** | 8214 |
| NC_005363 | *Bdellovibrio bacteriovorus* | **B** | 7170 |
| NC_002927 | *Bordetella bronchiseptica* | **B** | 9993 |
| NC_001318 | *Borrelia burgdorferi* | **B** | 3279 |
| NC_004463 | *Bradyrhizobium japonicum* | **B** | 16669 |
| NC_003317 | *Brucella melitensis* | **B** | 6398 |
| NC_002528 | *Buchnera aphidicola* | **B** | 1148 |
| NC_005061 | *Candidatus Blochmannia* | **B** | 1295 |
| NC_002696 | *Caulobacter vibrioides* | **B** | 7476 |
| NC_002620 | *Chlamydia muridarum* | **B** | 3080 |
| NC_002932 | *Chlorobaculum tepidum* | **B** | 4504 |
| NC_005085 | *Chromobacterium violaceum* | **B** | 8814 |
| NC_001263 | *Deinococcus radiodurans* | **B** | 6299 |
| NC_000913 | *Escherichia coli K-12* | **B** | 9412 |
| NC_000907 | *Haemophilus influenzae* | **B** | 3410 |
| NC_000915 | *Helicobacter pylori* | **B** | 3155 |
| NC_002945 | *Mycobacterium bovis* | **B** | 7840 |
| NC_000962 | *Mycoplasma tuberculosis* | **B** | 8082 |
| NC_000908 | *Mycoplasma genitalium* | **B** | 964 |
| NC_003112 | *Neisseria meningitidis* | **B** | 4107 |
| NC_004193 | *Oceanobacillus iheyensis* | **B** | 6996 |
| NC_005042 | *Prochlorococcus marinus* | **B** | 3764 |
| NC_002516 | *Pseudomonas aeruginosa* | **B** | 11152 |
| NC_003198 | *Salmonella enterica* | **B** | 5129 |
| NC_004347 | *Shewanella oneidensis* | **B** | 9254 |
| NC_004741 | *Shigella flexneri* | **B** | 8136 |
| NC_002758 | *Staphylococcus aureus* | **B** | 6016 |
| NC_003098 | *Streptococcus pneumoniae* | **B** | 4439 |
| NC_004113 | *Thermosynechococcus elongatus* | **B** | 4989 |
| NC_000919 | *Treponema pallidum* | **B** | 2067 |
| NC_002162 | *Ureaplasma urealyticum* | **B** | 1225 |
| NC_002505 | *Vibrio cholerae* | **B** | 7663 |
| NC_005090 | *Wolinella succinogenes* | **B** | 2044 |
| NC_003919 | *Xanthomonas axonopodis* | **B** | 8854 |
| NC_002488 | *Xylella fastidiosa* | **B** | 5664 |
| NC_004088 | *Yersinia pestis* | **B** | 8877 |
| NC_2607 | *Halobacterium salinarum* | **A** | 7667 |
| NC_003901 | *Methanosarcina mazei* | **A** | 6742 |

## Table S3

| **Accession number** | **Name** | **Domain of Life** | **Size** |
|---|---|---|---|
| NC_000854 | *Aeropyrum pernix* | A | 4535 |
| NC_000917 | *Archaeoglobus fulgidus* | A | 4829 |
| NC_000909 | *Methanococcus jannaschii* | A | 3557 |
| NC_003551 | *Methanopyrus kandleri* | A | 3374 |
| NC_005213 | *Nanoarchaeum equitans* | A | 1072 |
| NC_000868 | *Pyrococcus abyssi* | A | 1869 |
| NC_003413 | *Pyrococcus furiosus* | A | 4202 |
| NC_000961 | *Pyrococcus horikoshii* | A | 3690 |
| NC_002754 | *Sulfolobus solfataricus* | A | 5971 |
| NC_003106 | *Sulfolobus tokodaii* | A | 5653 |
| NC_000918 | *Aquifex aeolicus* | B | 3122 |
| NC_000853 | *Thermotoga maritima* | B | 3726 |

## Table S4

| **Amino acid residue** | ***EC*(%)** | ***PF*(%)** | **Exp in *PF*** | **Obs in *PF*** | **Diff in σ (p-value)** |
|---|---|---|---|---|---|
| **ARG** | 5.53 | 5.34 | 32469±525 | 31353 | $6.4(8\cdot 10^{-11})$ |
| **LYS** | 4.41 | 8.1 | 25893±471 | 47559 | $138(<10^{-14})$ |
| **ASP** | 5.1 | 4.36 | 29945±504 | 25600 | $26(<10^{-14})$ |
| **GLU** | 5.76 | 8.9 | 33820±534 | 52257 | $104(<10^{-14})$ |

**Figure S1**

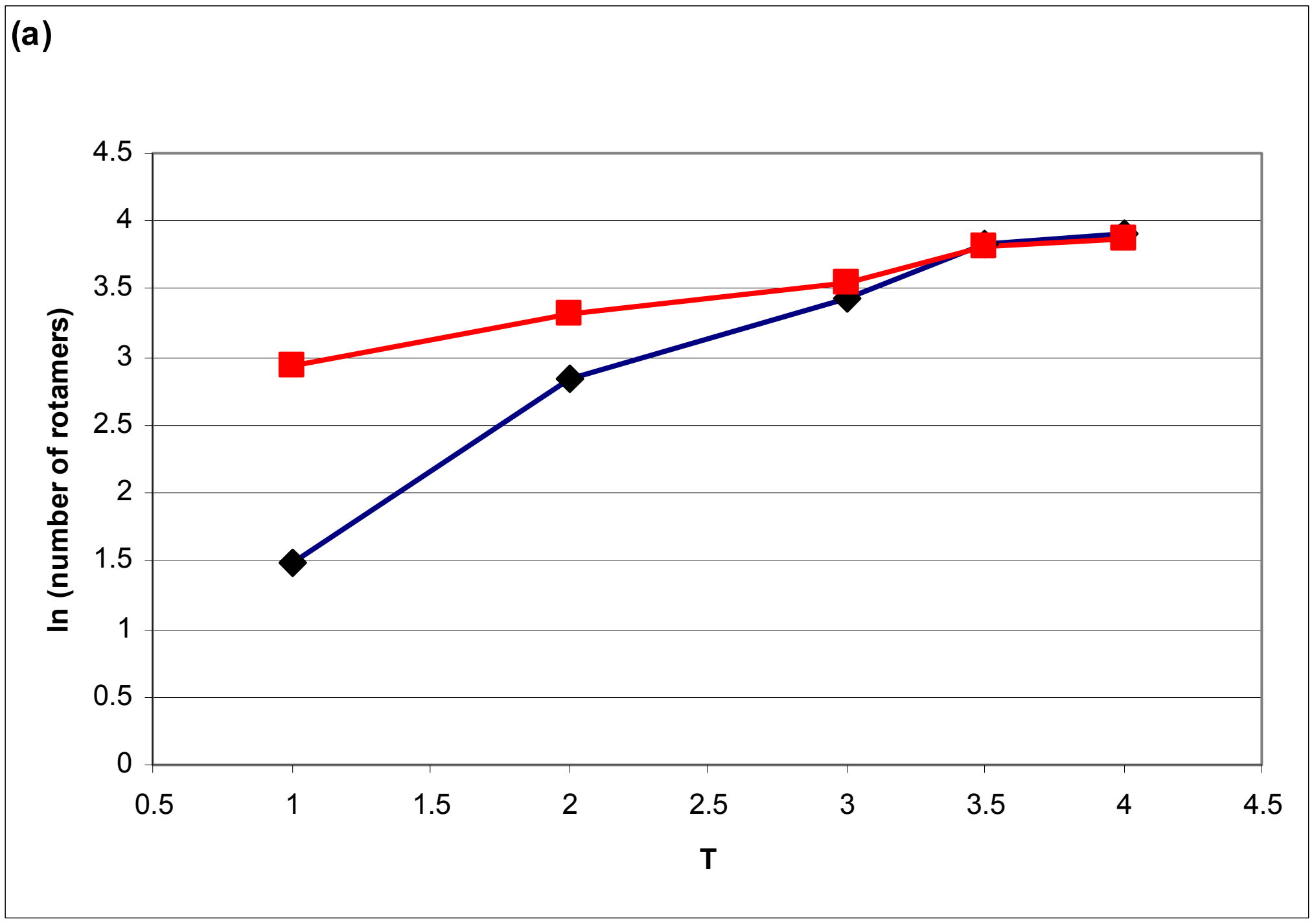

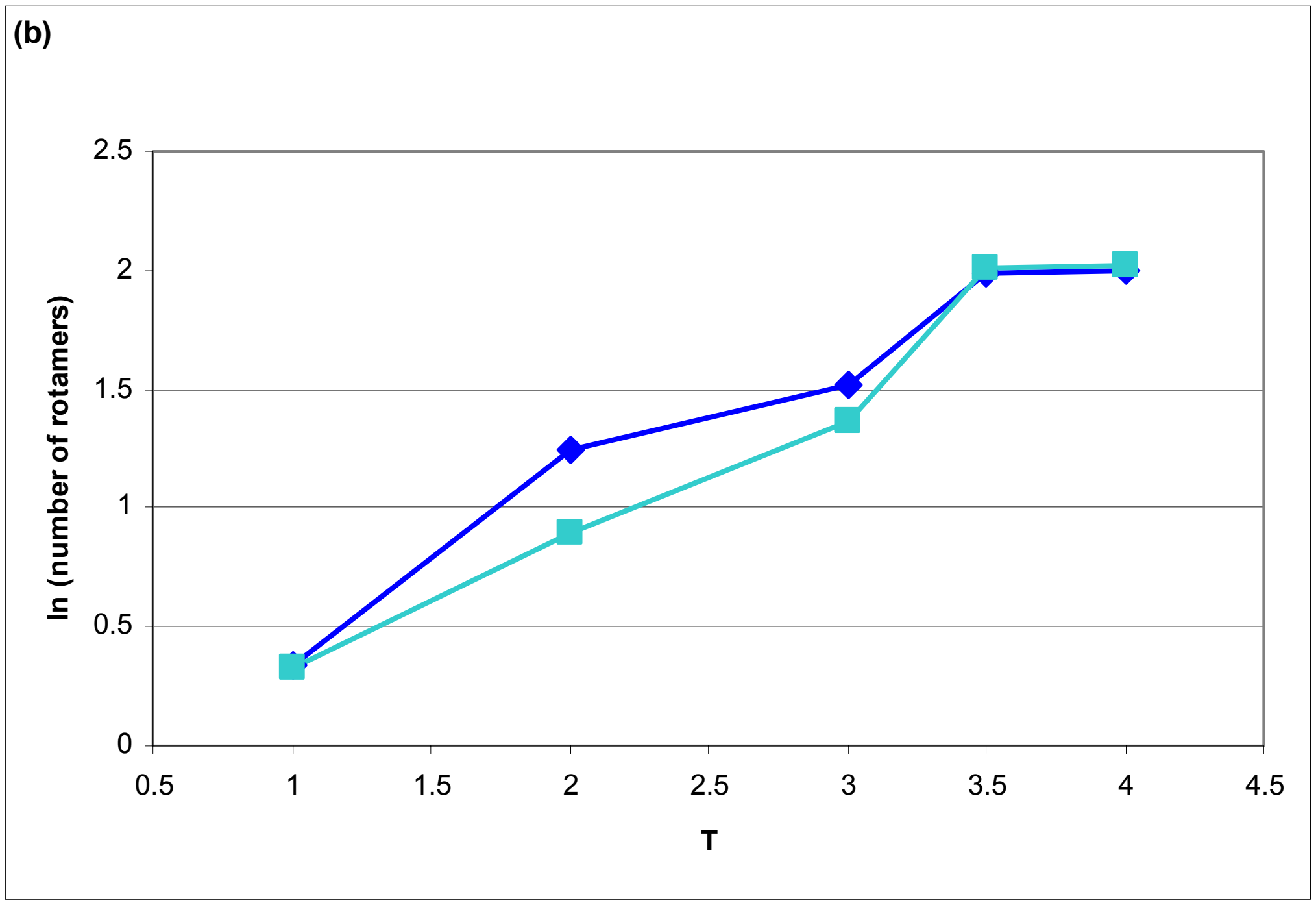

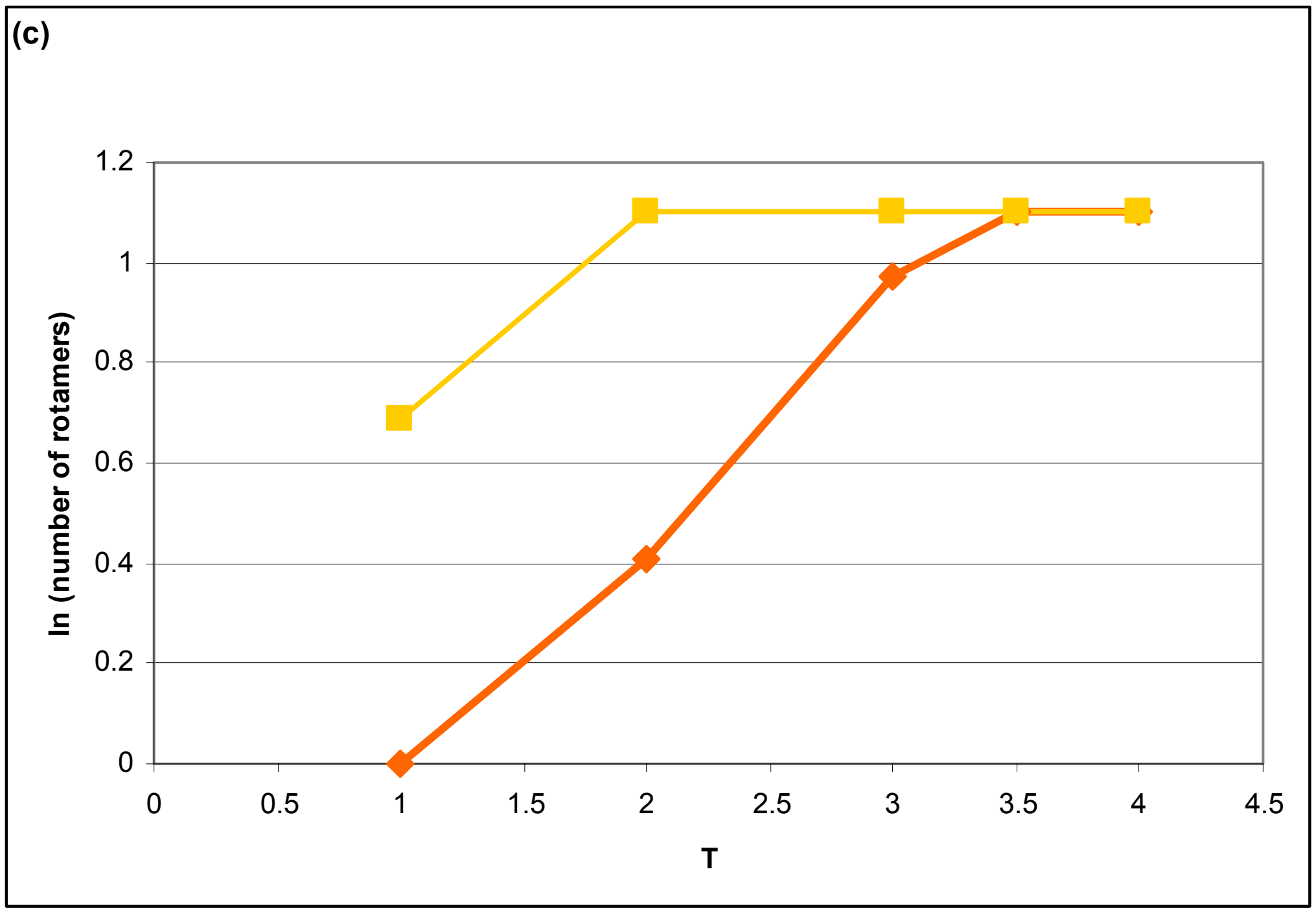

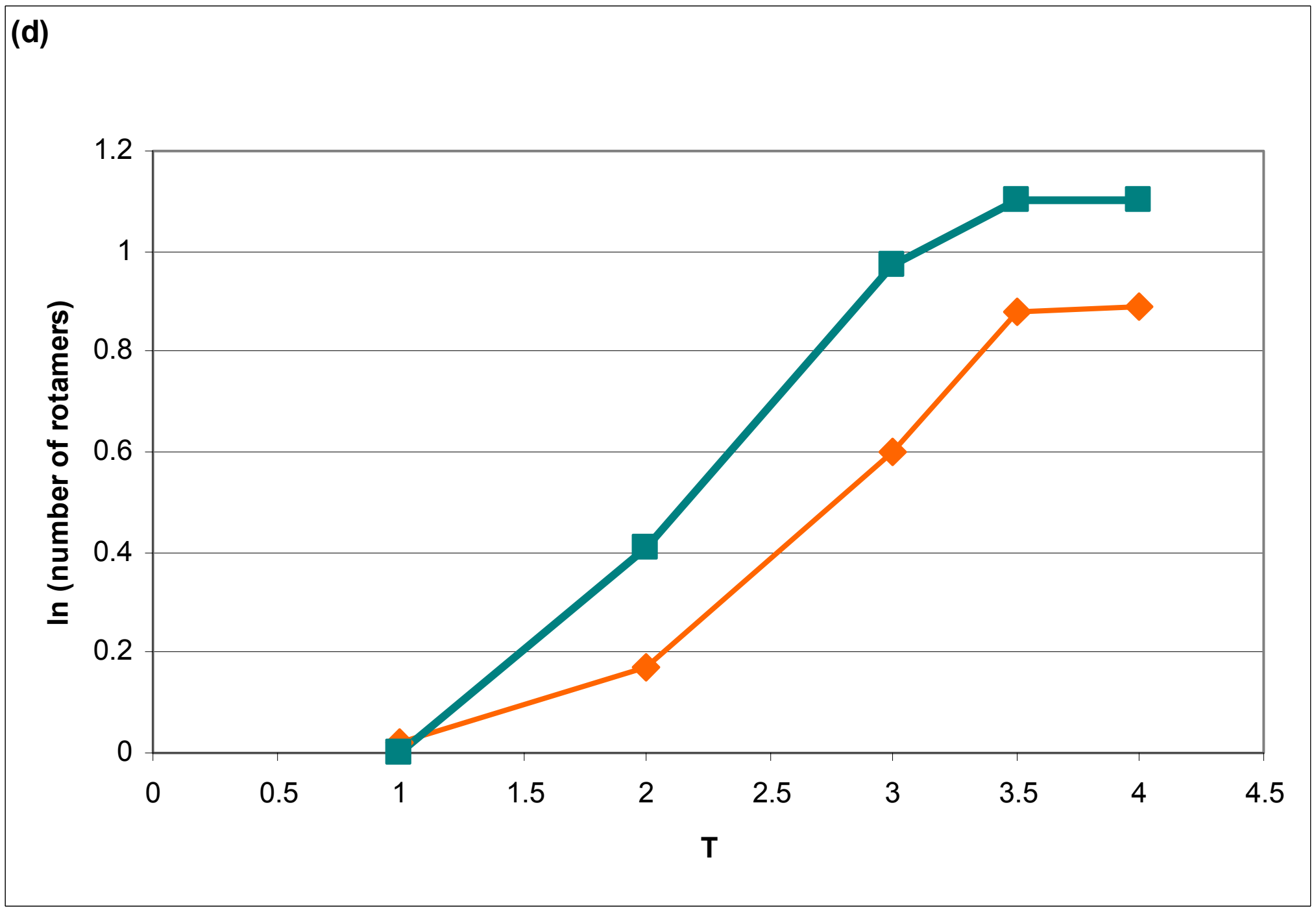

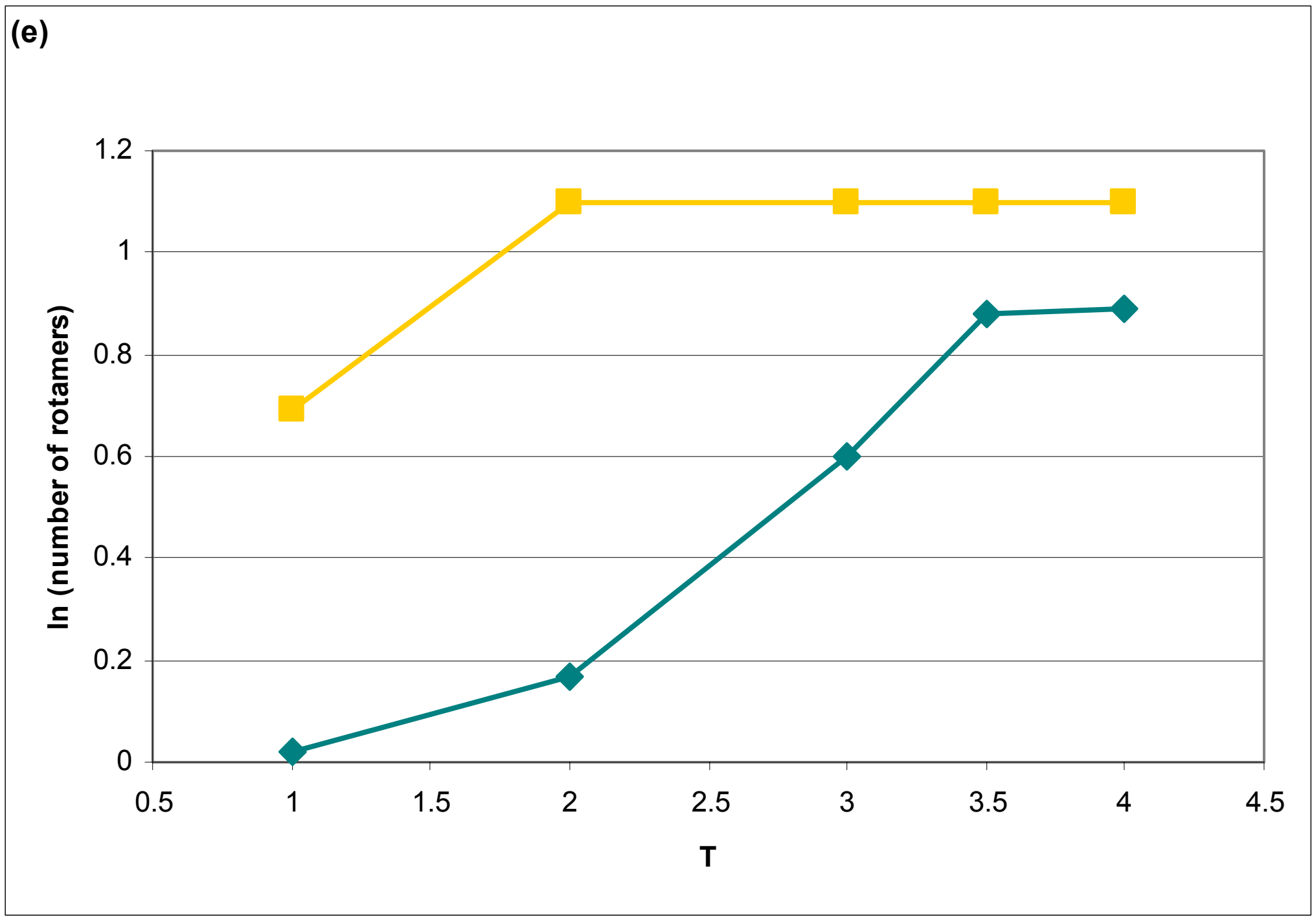

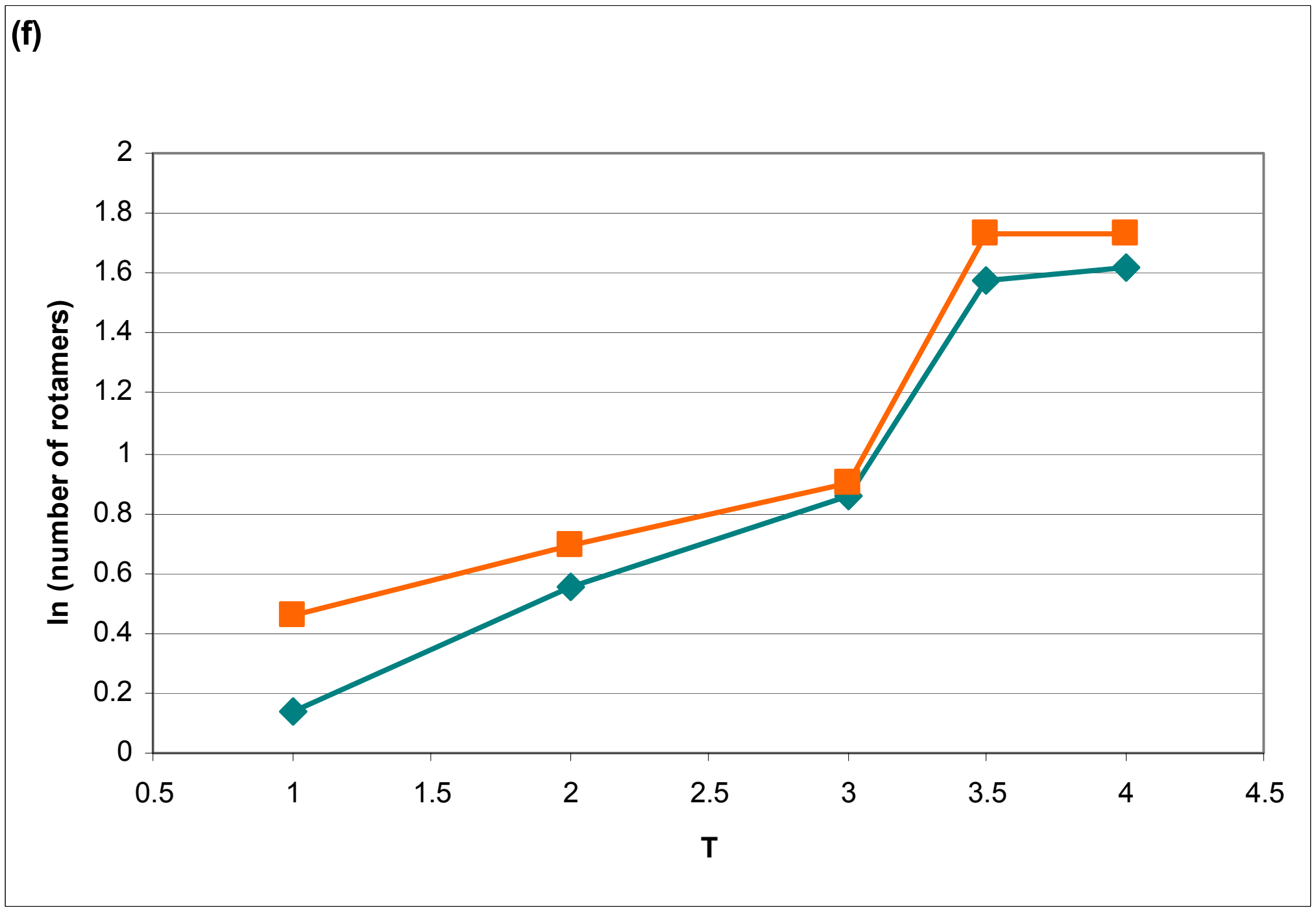

**Figure S2**

**(a)**

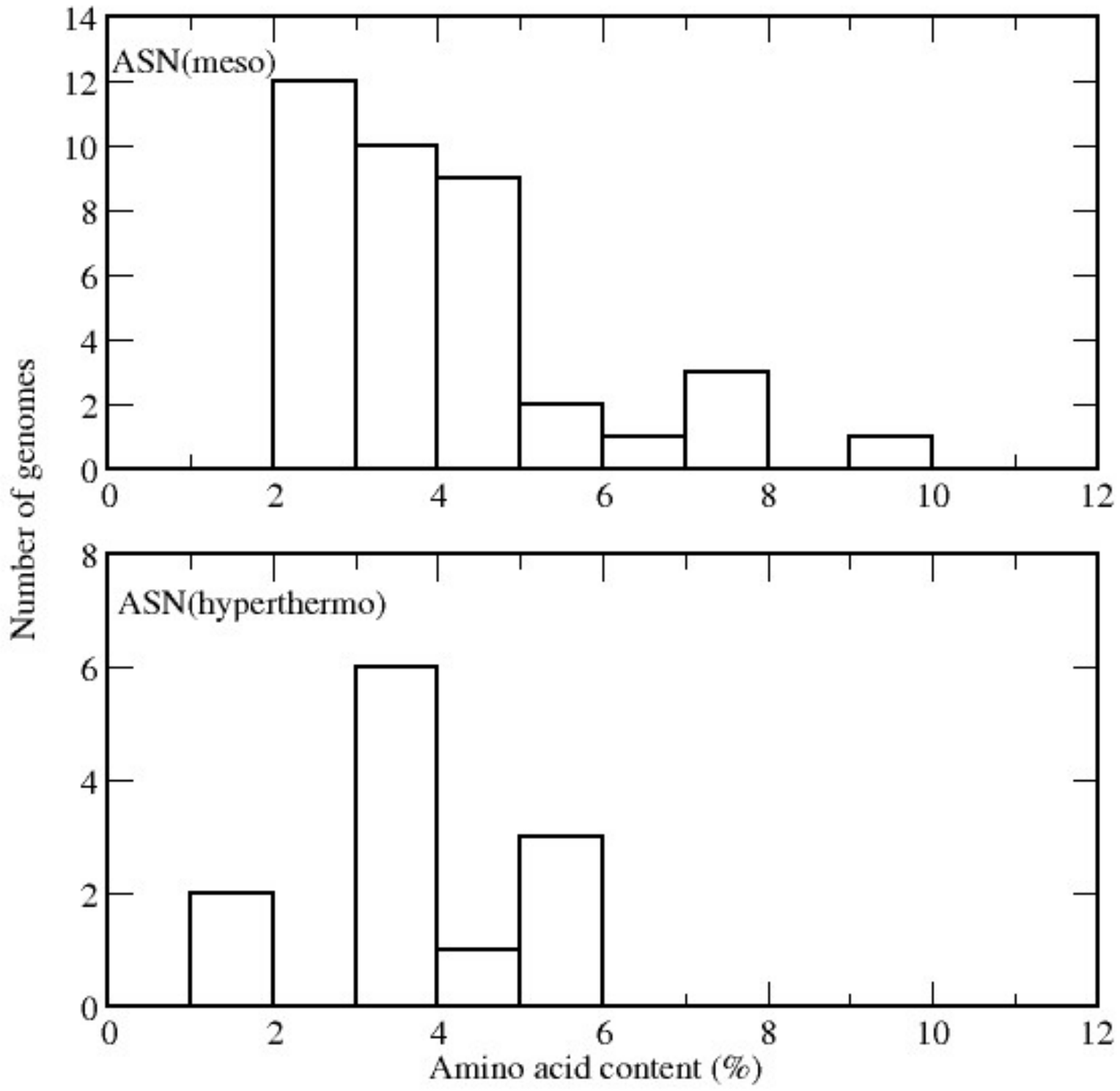

ASN(meso)
ASN(hyperthermo)
Number of genomes
Amino acid content (%)

**(b)**

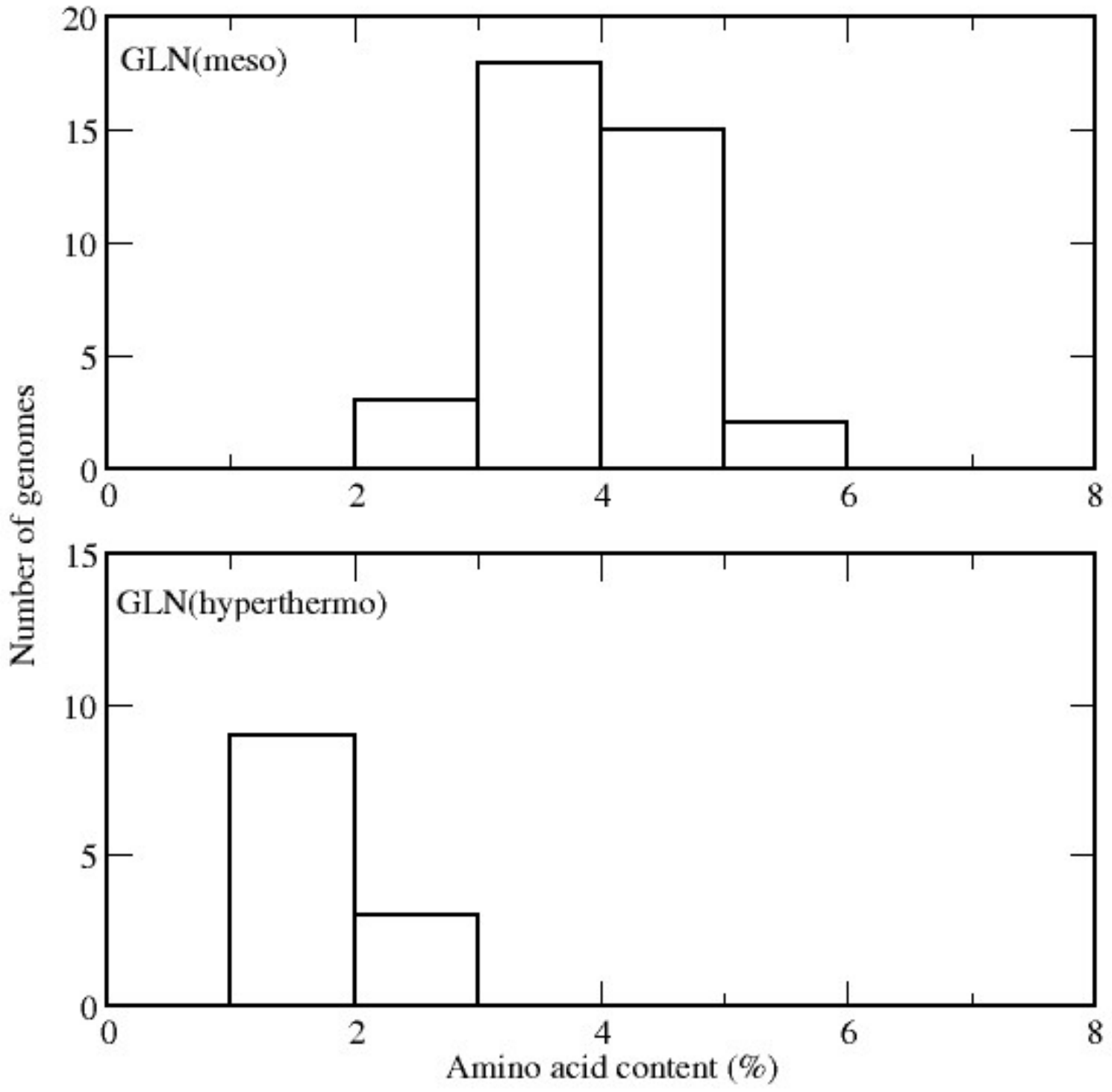

GLN(meso)
GLN(hyperthermo)
Number of genomes
Amino acid content (%)

**(c)**

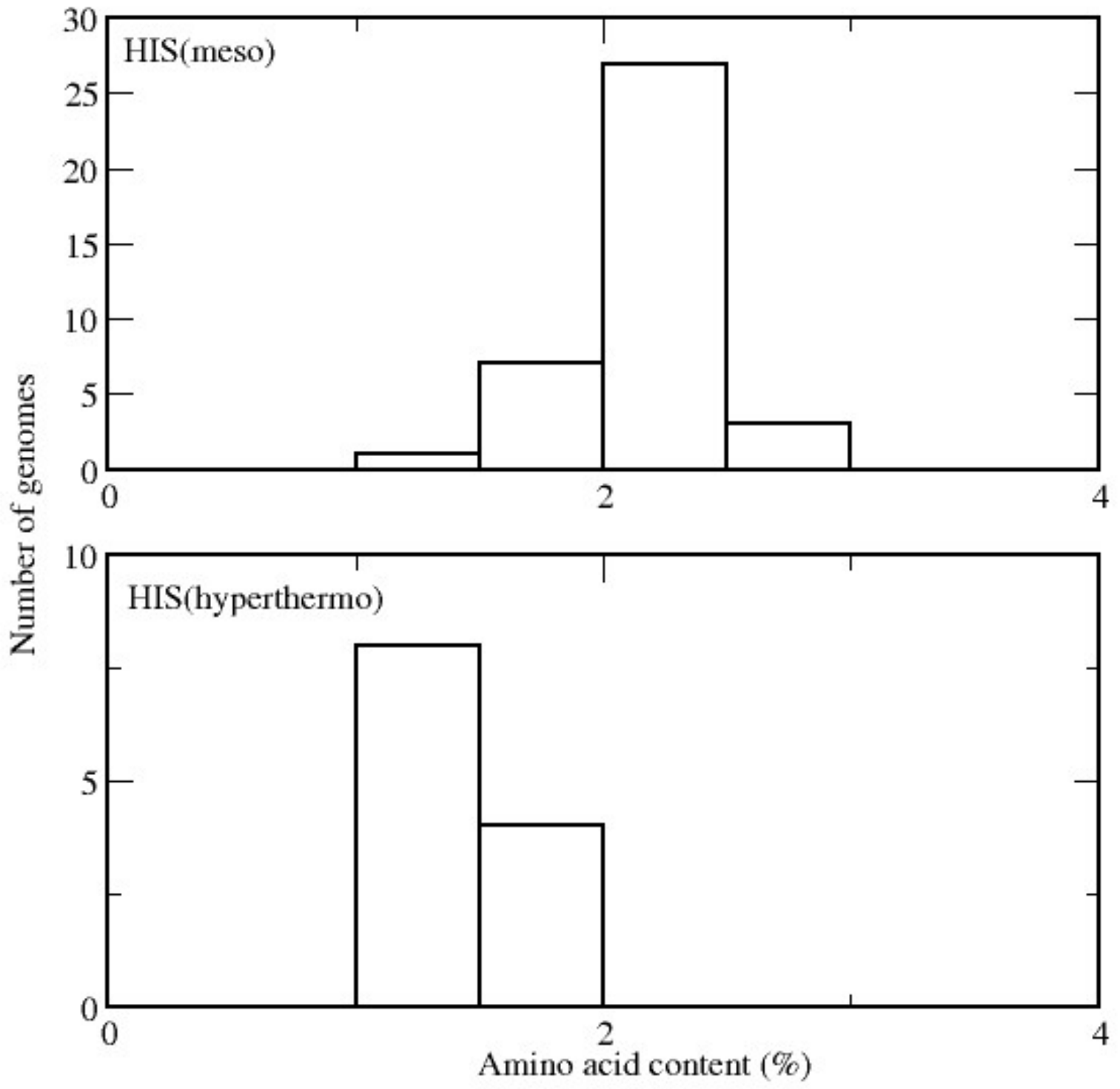

HIS(meso)
HIS(hyperthermo)
Number of genomes
Amino acid content (%)

**(d)**

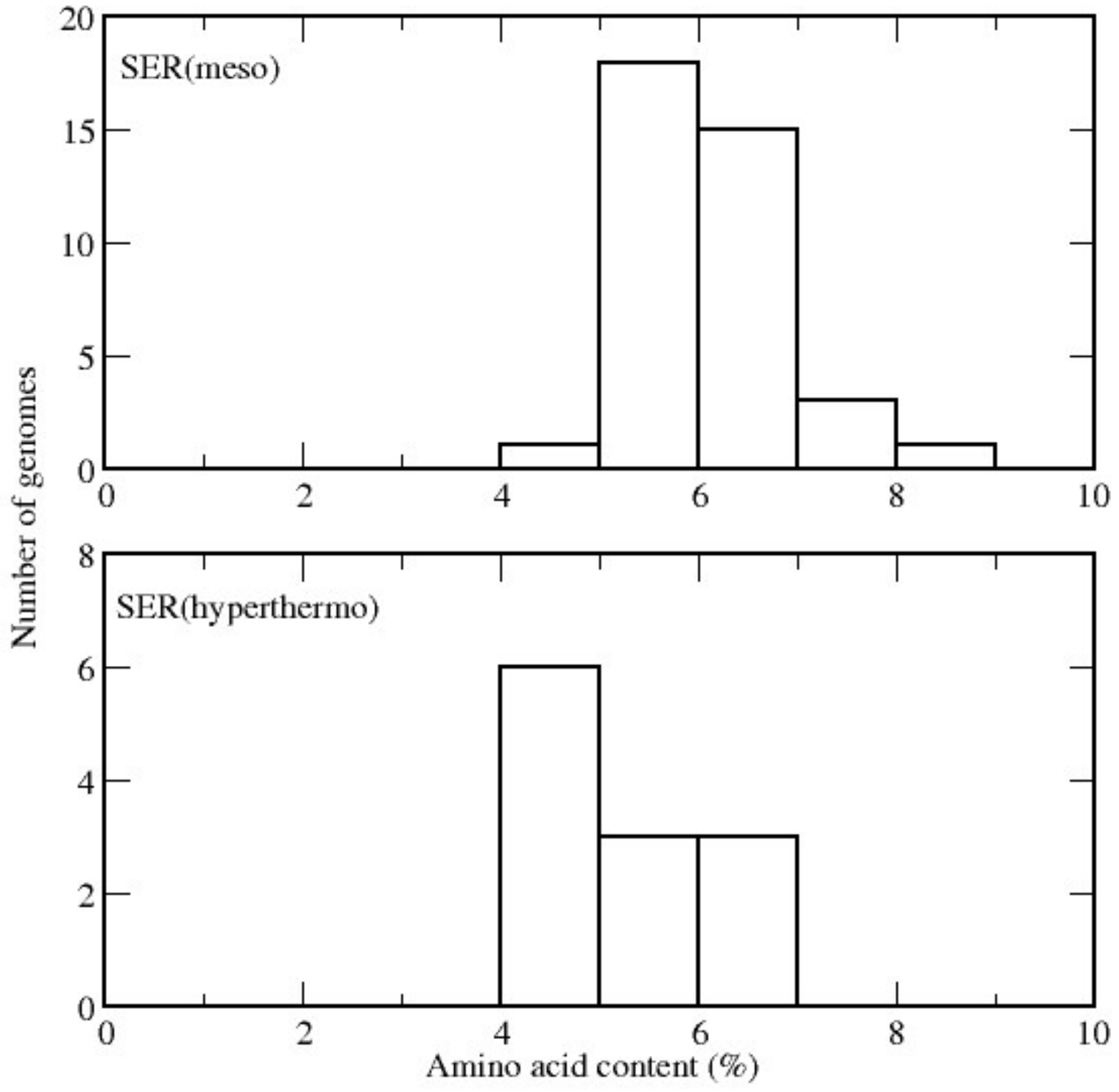

SER(meso)
SER(hyperthermo)
Number of genomes
Amino acid content (%)

**(e)**

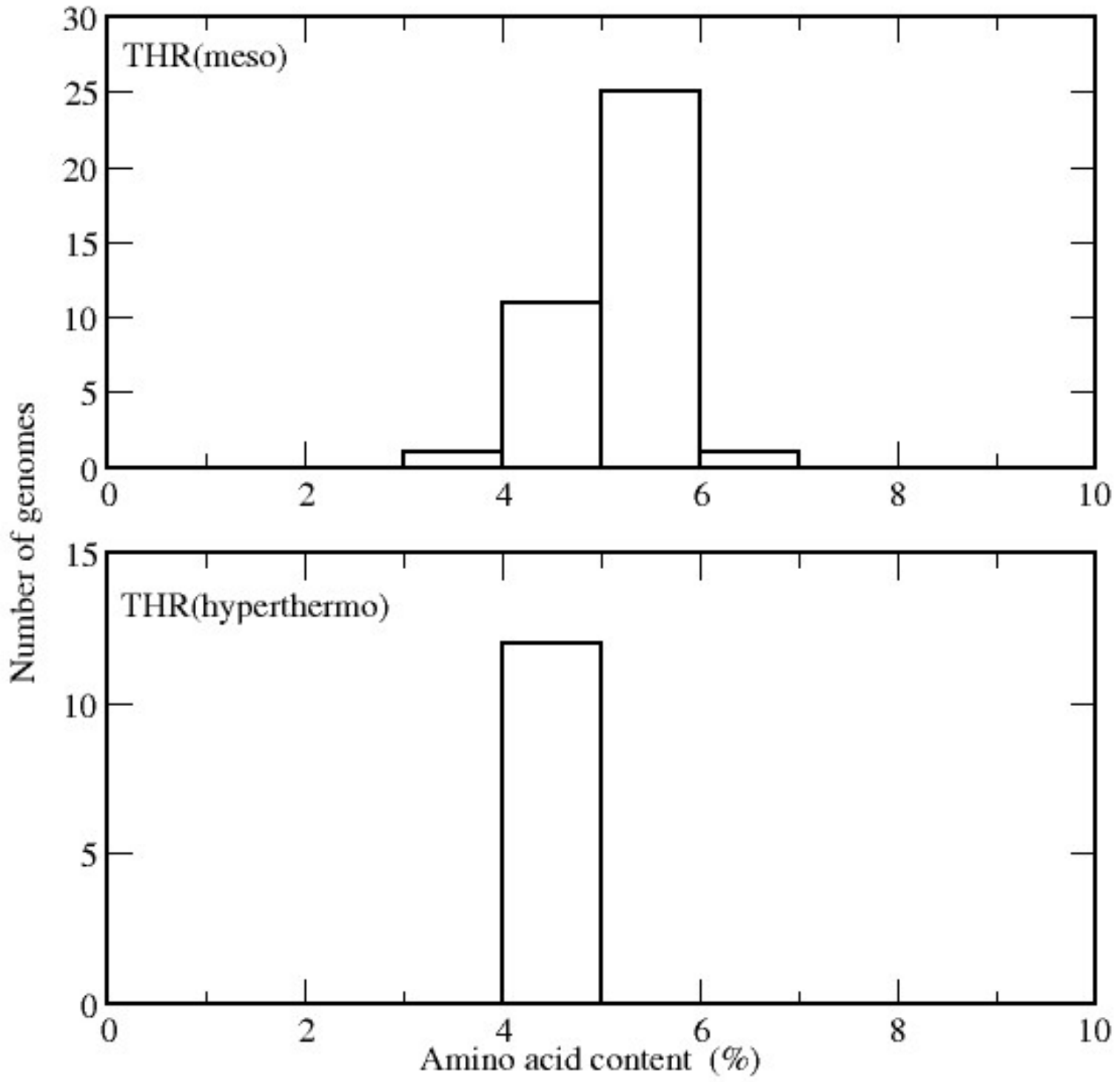

**(f)**

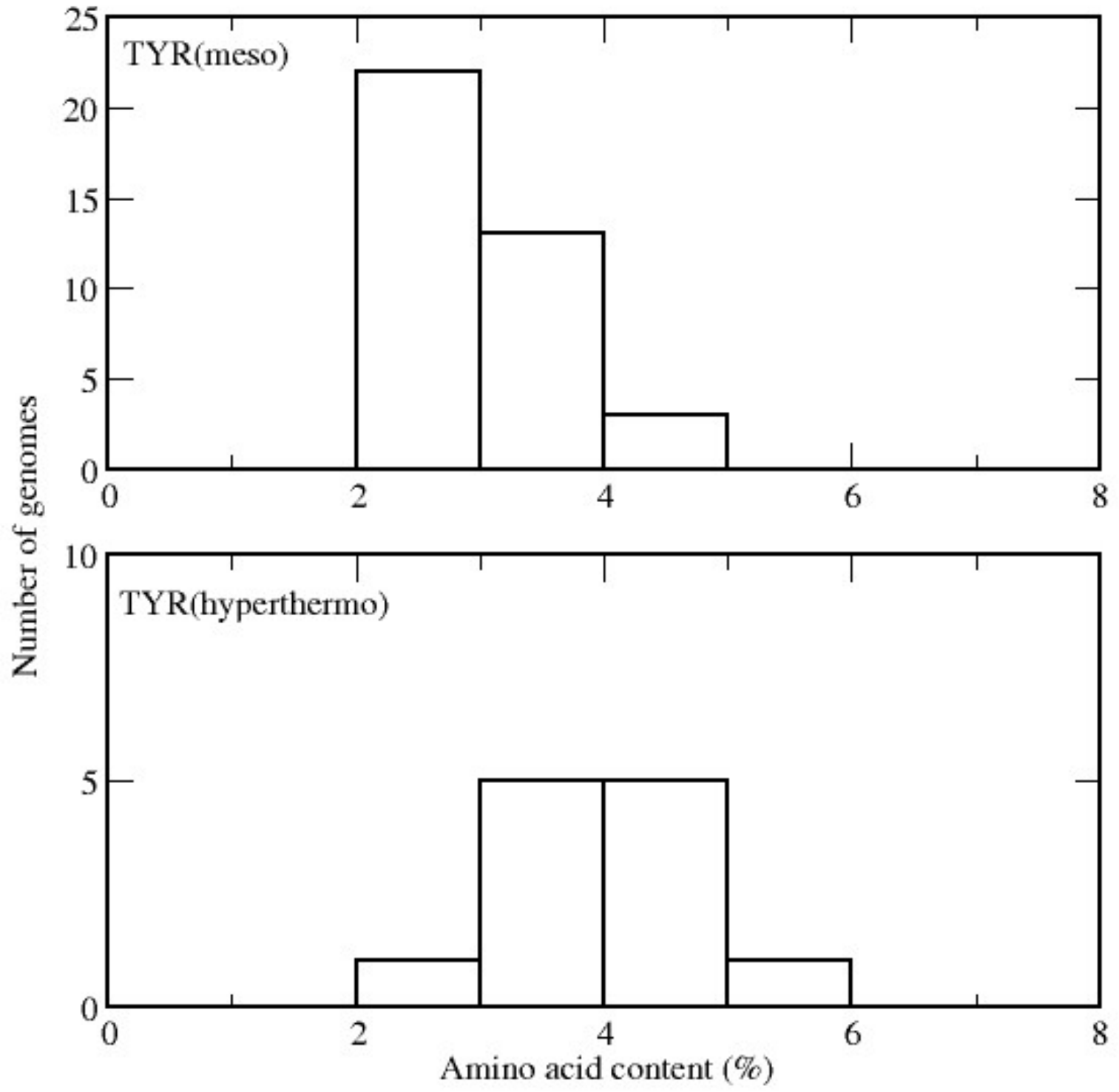